\documentclass[aps,pre,twocolumn,superscriptaddress,showpacs]{revtex4-1}
\usepackage{graphicx}% Include figure files
\usepackage{dcolumn}% Align table columns on decimal point
\usepackage{bm}% bold math

\usepackage{natbib}
\usepackage[
text={7in,9.5in},centering,
]{geometry}

\usepackage{epsfig}
\usepackage{amsmath}
\usepackage{amssymb}
\usepackage{textcomp}

\begin{document}

%\preprint{}

\title{Sensitivity of directed networks to the addition and pruning of edges and vertices.
}


\affiliation{Department of Physics $\&$ I3N, University of Aveiro, 3810-193 Aveiro, Portugal}
\author{A. V. Goltsev}
\affiliation{Department of Physics $\&$ I3N, University of Aveiro, 3810-193 Aveiro, Portugal}
\affiliation{A.F. Ioffe Physico-Technical Institue, 194021 St. Petersburg, Russia}
\author{G. Tim\'{a}r}
\affiliation{Department of Physics $\&$ I3N, University of Aveiro, 3810-193 Aveiro, Portugal}
\author{J. F. F. Mendes}
\affiliation{Department of Physics $\&$ I3N, University of Aveiro, 3810-193 Aveiro, Portugal}

%\thanks{}


%\date{\today}

\begin{abstract}
We study the sensitivity of directed complex networks to the addition and pruning of edges and vertices and introduce the susceptibility, which quantifies this sensitivity. We show that topologically different parts of a directed network have different sensitivity to the addition and pruning of edges and vertices and, therefore, they are characterized by different susceptibilities. These susceptibilities diverge at the critical point of the directed percolation transition, signaling the appearance (or disappearance) of the giant strongly connected component in the infinite size limit. We demonstrate this behavior in randomly damaged real and synthetic directed complex networks, such as the World Wide Web,  Twitter, the \emph{Caenorhabditis elegans} neural network, directed Erd\H os-R\'enyi graphs, and others. We reveal a non-monotonous dependence of the sensitivity  to random pruning of edges or vertices in the case of \emph{Caenorhabditis elegans}  and Twitter that manifests  specific structural peculiarities of these networks. We propose the measurements of the  susceptibilities during the addition or pruning of edges and vertices as a new method for studying structural peculiarities of directed networks.
\end{abstract}

   %Example=paper!!
%\keywords{}
\maketitle

% -------------------------- Introduction ---------------------------------------------------------------------------------
\section{Introduction}
\label{introduction}

Many real-world complex systems can be represented by directed complex networks. In this kind of network every edge between two interacting subjects
%is directed, i.e., it goes in only one direction.
can be directed, i.e., it goes in only one direction, and bidirected, i.e., it goes in both directions.
Well-known examples are the World Wide Web (WWW) \cite{broder2000graph}, neuronal and metabolic networks \cite{sporns2004organization,barabasi2004network}, gene regulatory networks \cite{lee2002transcriptional,liu2015regnetwork}, social networks, such as Twitter \cite{kwak2010twitter}, the control network of transnational corporations \cite{vitali2011network}, and many other complex systems \cite{newman2003structure,dorogovtsev2002evolution}.
A common feature of these complex systems is that they are developing by means of the addition of directed edges and vertices while diseases, injuries, random or targeted damages lead to the degradation of the network structure due to pruning of edges or vertices. For example, in the brain, neurogenesis (the creation of new nerve cells) and the formation of new synaptic connections, as well as the opposite process -- pruning of synapses or neurons, are important mechanisms for the brain development \cite{sporns2004organization,bullmore2012economy}, learning and memory \cite{chklovskii2004cortical,brunel2016cortical}, and sex-specific circuit development \cite{oren2016sex,portman2016neurobiology}, and for other brain functions. %\cite{sporns2004organization,bullmore2012economy,chklovskii2004cortical,brunel2016cortical,oren2016sex,portman2016neurobiology}.
In contrast to neurogenesis, neurodegenerative diseases are accompanied by the loss of neurons or synapses and degradation of the brain networks \cite{stephan2006synaptic,palop2010amyloid,benes2001density}. Twitter, a social network, is developing in a similar way, growing due to addition of new users and the formation of new connections \cite{kwak2010twitter}.
The structure of these directed networks plays a very important role in their functioning. The structure of a directed network is  much more subtle and richer than the structure of its undirected version \cite{broder2000graph,newman2001random,dorogovtsev2001giant,timar2017mapping}.
In general, a directed graph consists of the giant strongly connected component $G_S$, which is a central core of the  network, the sets $IN$ and $OUT$  of vertices playing the role of the incoming and outgoing terminals for $G_S$.
%from which $G_S$ can be reached by following edges forwards or backwards, respectively.
There are also finite directed components $F$ (tendrils, tubes, and disconnected finite clusters).
If in the initial state a network consists of isolated vertices and disconnected finite clusters, then the addition of  directed edges leads to the appearance of $G_S$ at the critical point of the directed percolation transition.
Edges can be added at random, as in the case of ordinary percolation, or by exploiting some optimization principle, for example, the Achlioptas process, as in the case of explosive percolation \cite{achlioptas2009explosive,squires2013weakly}.
In networks subjected to pruning of edges or vertices, $G_S$ disappears at the  critical point below which the network disintegrates into a set of finite directed components and disconnected clusters.
Taking into account the crucial role of $G_S$ in dynamics, functioning, and the propagation of information in directed complex networks, it is very important to develop methods that quantify the sensitivity of the networks to structural changes and signal the approaching of the critical transition.
In physics, the sensitivity of a system to an applied field is quantified by the susceptibility, which characterizes the sensitivity of the order parameter to an applied field (see, for example, \cite{stanley1971introduction}).
The susceptibility diverges in the infinite size limit when a system under consideration approaches the critical point of a continuous phase transition.
Thus the susceptibility not only quantifies the response of a system to an applied field but also signals the approaching of the phase transition.
The generalized susceptibility was already discussed in the case of ordinary percolation  \cite{stauffer1994introduction} and explosive percolation \cite{da2010explosive,da2014solution} in  undirected complex networks. In the case of directed complex networks, the susceptibility was recently introduced in \cite{timar2017mapping}. However, a relation between the susceptibility and the sensitivity of directed networks to
%structural perturbations
the addition and pruning of edges or vertices was not yet studied. Since different network parts ($G_S$, $IN$, $OUT$, and $F$) have different topological properties, one can assume that these parts also differ in the sensitivity to the addition and pruning of edges or vertices and, therefore, they must be characterized by different susceptibilities. This problem has not yet been discussed in the literature.

In this paper, we study the response of directed networks to the addition and pruning of edges and vertices. We show that topologically different network parts ($IN$, $OUT$, $G_S$, and $F$) have different sensitivities to this kind of impact on the networks. These sensitivities are quantified by the different susceptibilities, which can be found by use of the two-point connectivity function characterizing whether any two vertices are connected by a directed path or not. Alternatively, we find the susceptibilities by analyzing statistics of individual  out-components and in-components of vertices in $IN$, $OUT$, $G_S$, and $F$. We use these local characteristics for quantifying the response of the giant component $G_S$, $IN$, $OUT$, and the finite network components $F$ to the addition and pruning of edges or vertices. Using the generating function method, we find analytically the susceptibilities of uncorrelated random directed complex networks  in the infinite size limit and demonstrate that the susceptibilities
%achieves a maximum (
diverge at the critical point, signaling the percolation transition in these networks.
Finally, we study numerically the susceptibilities of some real and synthetic directed complex networks subjected to random damage and show that these susceptibilities characterize the sensitivity of even small directed networks to damage and depend strongly on structural peculiarities of the directed networks.

\section{Structure of directed networks}
\label{structure}
In order to find the response of a directed network to the addition or removal of edges or vertices, we first need to know its structure.
In this section we describe the common structural properties of directed graphs revealed in  \cite{broder2000graph,newman2001random,dorogovtsev2001giant,schwartz2002percolation,boguna2005generalized,bianconi2008local,timar2017mapping}.

\begin{figure}[htpb!]
\centering
\includegraphics[width=6cm,angle=0.]{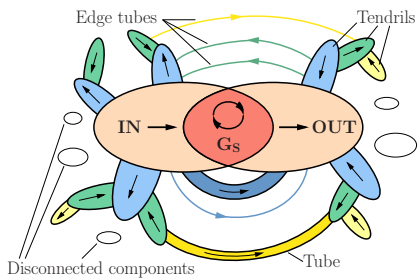}
\caption{(Color online) Schematic view of the structure of a directed
network. There are the giant strongly connected component $G_S$ (the core of the network), the sets $IN$ and $OUT$, %in-component ($G_{in}$) and out-component  ($G_{out}$) of $G_S$,
the finite directed components $F$ [tendrils and tubes shown as domains of different colors, and disconnected finite clusters (open ovals)].
The union of these parts is the giant weakly connected component $G_W$.
%Furthermore, there are disconnected finite clusters (open ovals).
%Different tendril layers are shown by different colors.
Tendrils belonging to the same layer have the same color.
%Edge-tubes are shown as blue and green directed edges.
Edge tubes of different layers are shown in corresponding colors. Only three tendril layers are shown.
%here but, in general, there can be any number of these layers.
}
\label{fig:structure}
\end{figure}

Directed networks have a hierarchical organization \cite{broder2000graph,newman2001random,dorogovtsev2001giant,timar2017mapping} and can be partitioned into topologically different parts (see Fig. \ref{fig:structure}):
(i) the giant strongly connected component ($G_S$), which is a central core of the directed network, (ii) sets of vertices called $IN$ and $OUT$  that are connected to $G_S$, (iii) hierarchically organized finite directed components (tendrils and tubes) that are only connected to $IN$ and $OUT$ but not to $G_S$, and (iv) disconnected finite clusters.
These parts have different topological properties. The definitions of these network parts were given in \cite{broder2000graph,newman2001random,dorogovtsev2001giant,timar2017mapping}. Let us briefly review them. The giant strongly connected component $G_S$
%(throughout this paper, we also use the notation $G_S$ for the brevity of $G_S$),
is a subgraph in which every vertex can be reached from every other vertex by following directed edges.
$OUT$ is a set of vertices that can be reached from the $G_S$ by following directed edges, but from which it is not possible to reach the $G_S$.
$IN$ is a set of vertices from which the strongly connected
component $G_S$ can be reached by following directed edges, but which can not be reached from the strongly connected component by following directed edges. Note that $IN$ and $OUT$ may appear only when $G_S$ appears. In some real directed complex networks, such as the neural network of \emph{Caenorhabditis elegans} (\emph{C. elegans}) \cite{jarrell2012connectome}, the giant strongly connected component $G_S$ includes almost all vertices of the considered network (492 vertices among 495 vertices in \emph{C. elegans}, see Sec. \ref{simulations} for more details),
%i.e.,   $G_S = \mathcal{G}$,
as a condition of its normal functioning  \cite{timar2017mapping}.

The union of the sets $G_S$ and $OUT$ is the giant out-component $G_{out}$ \cite{dorogovtsev2001giant}, i.e.,
%$G_{out}= G_S \cup OUT$.
\begin{equation}
 G_{out}= G_S \cup OUT.
  \label{eq: 1OUT}
\end{equation}
In turn, the union of $G_S$ and $IN$ is the giant in-component $G_{in}$ \cite{dorogovtsev2001giant}, i.e.,
%$G_{in}= G_S \cup OUT$.
%\cite{broder2000graph,newman2001random,dorogovtsev2001giant}
\begin{equation}
 G_{in}= G_S \cup IN.
  \label{eq: 1IN}
\end{equation}
The remaining part ($F$) of the graph $\mathcal{G}$ is the union of finite components including finite directed components $T$ (tendrils and tubes)
and finite disconnected clusters $C$, i.e.,
%$F=T \cup C$.
\begin{equation}
 F = T \cup C =\mathcal{G} \setminus (G_{S} \cup IN \cup OUT).
  \label{eq: 1F}
\end{equation}
Tendrils and tubes in $T$ have a hierarchical, multilayer organization around $IN$ and $OUT$
\cite{timar2017mapping}. They can exist only when $IN$ and $OUT$ are present in the network.
The set of vertices $C$ is the union of all finite disconnected clusters $C_{\alpha}$, $\alpha =1, 2, 3, \dots$, i.e., ($C=C_1 \cup C_2 \cup C_3 \dots$).

The giant weakly connected component $G_W$ is the union of $G_S$, $IN$, $OUT$, tendrils, and tubes, i.e.,
\begin{equation}
 G_W=G_{S} \cup IN \cup OUT \cup T= \mathcal{G} \setminus C.
  \label{eq: 1w}
\end{equation}
If we neglect the edge directedness, we find that the subgraph  $G_W$ is the giant connected component of the undirected version of the graph $\mathcal{G}$. Note that $G_W$ can exist even if  $G_S$, $IN$, and $OUT$ are absent in the network.

We also introduce a subgraph $G_{IO}$ as the union
 \begin{equation}
 G_{IO} \equiv G_{S} \cup IN \cup OUT = \mathcal{G} \setminus F.
  \label{eq: 1op}
\end{equation}
%i.e., the $G_{IO}$  is a set of vertices reachable by following edges both backwards and forwards from any vertex in the $G_S$.
We suggest that the size of $G_{IO}$ is the order parameter for the percolation transition in directed networks \cite{timar2017mapping}. Note that in undirected networks, the size  of the giant connected component (the undirected version of $G_W $) is the order parameter for the ordinary percolation.
There are similarities between the topological structure of the order parameters $G_{IO}$ and the giant connected component (the undirected version of $G_{W}$) of an undirected network.
The latter consists of the 2-core, i.e., the largest subgraph whose vertices have degree at least 2, and finite branches attached to this 2-core \cite{dorogovtsev2008critical}. In directed networks, $G_{IO}$ includes the subgraph $G_{S}$, whose vertices also have degree at least two (at least one incoming edge and at least one outgoing edge with vertices within the subgraph $G_{S}$). Vertices in $IN$ and $OUT$ form directed incoming and outgoing branches, respectively, attached to the $G_{S}$.
% (as it will be shown below).

In the general case, any directed graph $\mathcal{G}$ can be written as the following union:
\begin{equation}
 \mathcal{G}= G_{IO} \cup F= G_W \cup C = G_{S} \cup IN \cup OUT \cup T \cup C.
  \label{eq: 1}
\end{equation}
If there is no giant component $G_S$, the graph $\mathcal{G}$ consists of only finite directed components and finite disconnected clusters, i.e., $\mathcal{G}=F$.

If a directed network consists of only disconnected finite clusters then the addition of new directed edges results at first in the ordinary percolation transition into state in which the  undirected version of this directed network has a nonzero giant connected component.
Then, adding more directed edges, the network undergoes the directed percolation transition into a state with a nonzero giant strongly connected component $G_{S}$.

Let us characterize the network parts, i.e., $G_S$, $IN$, $OUT$, and the finite directed components, using the sizes of the individual in- and out-components of vertices $i$  \cite{newman2001random,dorogovtsev2001giant,timar2017mapping}.
By definition, the in-component  and out-component of vertex  $i$ are the sets of vertices reachable by following edges either backwards or forwards from  $i$, respectively.
%The sizes, $s_{in}(i)$ and  $s_{out}(i)$,  are defined by Eqs. (\ref{eq: 3}) and (\ref{eq: 4}).
If vertex $i$ belongs to $F= T \cup C$
%tendrils or  disconnected finite clusters (i.e., $i \in T \cup C$)
then it has finite in- and out-components, i.e., their  sizes are of the order of $O(1)$ [i.e., $s_{in}(i), s_{out}(i)\propto O(1)] $. Vertices belonging to  $G_S$ have equal sizes of in-components and equal sizes of out-components, namely, $s_{in}(i)=N(G_{in})-1$ and $s_{out}(i)=N(G_{out})-1$ for any $i \in G_S$. These individual components are giant, i.e., $s_{in}(i), s_{out}(i)\propto O(N) $. If $i \in IN$ then $s_{in}(i) \propto O(1)$ while  $s_{out}(i) \propto O(N)$. If $i \in OUT$ then $s_{in}(i) \propto O(N)$ while  $s_{out}(i) \propto O(1)$. Note that in general $s_{in}(i)$, as well as $s_{out}(i)$, are different for different $i$ in both $IN$ and $OUT$.

In Secs. \ref{network response} - \ref{simulations} we will study how the addition and pruning of edges and vertices affect the components of directed networks and their sensitivity to damage both below and above the directed percolation  transition.

\section{Two-point connectivity function}
\label{corr function}

Let us consider an arbitrary directed graph $\mathcal{G}$ of size $N(\mathcal{G}) \equiv N$.
 %[in this paper we will also use the notation $|\mathcal{G}|$ for $N(\mathcal{G})$].
In order to characterize
%weather any two vertices $i$ and $j$  are connected by a directed path or not,
the connectivity of the graph, we introduce  a two-point connectivity function $C(i \rightarrow j)$ of vertices $i$ and $j$ as follow: (i) $C(i \rightarrow i)=1$; (ii) $C(i \rightarrow j)=1$ if there is a directed path from $i$ to $j$
%by following edges forwards
(note that there can be more than one path).
%such that we start at $i$ then move  along the edge direction and finally we end up at $j$.
Otherwise, $C(i \rightarrow j)=0$.
%This function characterizes the connectivity of the directed network.
In general $C(i \rightarrow j)$ is asymmetric, $C(i \rightarrow j) \neq C(j \rightarrow i)$, since a directed path from $j$ to $i$ can be present while a directed path from $i$ to $j$ can be absent, and vice versa. The function $C(i \rightarrow j)$ is the generalization of the two-point correlation function of the one-state Potts model in undirected networks \cite{wu1982potts} to the case of directed networks. The two-point connectivity function $C(i \rightarrow j)$ is determined by the adjacency matrix $A_{ij}$. This relation can be written in the form
%but the functional form of this relation is unknown
\begin{equation}
  C(i \rightarrow j)= \Theta (\sum_{n=1}^{\infty} (A^{n})_{ij}),
  \label{eq: 2a}
\end{equation}
at $i\neq j$. The theta-function $\Theta(x)$ is 1 if $x>0$ and zero otherwise.
Let us also introduce the function,
\begin{equation}
  C(i,j)\equiv C(i \rightarrow j) + C(j \rightarrow i) - C(i \rightarrow j)C(j \rightarrow i),
  \label{eq: 2}
\end{equation}
which is symmetric, i.e., $C(i,j)=C(j,i)$. Furthermore, $C(i,i)=1$ and $C(i,j)=1$ if there is a %directed path from $i$ to $j$ either along or against the edge directions, or in both directions
directed path from $i$ to $j$ or from $j$ to $i$, or in both directions. Otherwise, $C(i,j)=0$.

Using the function $C(i \rightarrow j)$,
we can find the individual in- and out-components of every vertex $i$, which are defined as the
%As we defined in the previous section, in- and out components of vertex  $i$ are
sets of vertices reachable by following edges either backwards or forwards from  $i$, respectively. The sizes $s_{in}(i)$ and $s_{out}(i)$ of these components are
\begin{eqnarray}
  s_{in}(i)&=&\sum_{j \in \mathcal{G} \backslash i}C(j \rightarrow i),
  \label{eq: 3} \\
  s_{out}(i)&=&\sum_{j \in \mathcal{G}\backslash i}C(i \rightarrow j),
  \label{eq: 4} \\
   s_{t}(i)&=&\sum_{j \in \mathcal{G} \backslash  i}C(i,j),
  \label{eq: 5}
  \end{eqnarray}
where the sum is over all vertices $j \in \mathcal{G}$ except $i$. Here, $s_{t}(i)$ is the total number of vertices reachable by following edges  both backwards and forwards from $i$. Using Eqs. (\ref{eq: 3}) and (\ref{eq: 4}), we find that the mean sizes $\langle s_{in} \rangle $ and $\langle s_{out} \rangle $  of the individual in- and out-components of a randomly chosen vertex are equal to each other:
%\begin{equation}
%  \langle s_{in} \rangle \equiv \frac{1}{N(\mathcal{G})}\sum_{i \in \mathcal{G}} s_{in}(i) = \frac{1}{N(\mathcal{G})} \sum_{i \in \mathcal{G}} s_{out}(i) = \langle s_{out} \rangle.
%  \label{eq: 6}
%\end{equation}
%%
\begin{eqnarray}
  \langle s_{in} \rangle &\equiv& \frac{1}{N(\mathcal{G})}\sum_{i \in \mathcal{G}} s_{in}(i) =
  \frac{1}{N(\mathcal{G})}\sum_{i\neq j \in \mathcal{G}} C(j \rightarrow i)
  \nonumber \\
  &=&\frac{1}{N(\mathcal{G})} \sum_{j \in \mathcal{G}} s_{out}(j) = \langle s_{out} \rangle.
  \label{eq: 6}
\end{eqnarray}

In the general case we have $s_{t}(i) \leq s_{in}(i) + s_{out}(i)$ for vertices $i \in \mathcal{G}$ because there might be vertices that belong simultaneously to in- and out-components of vertex $i$ due to loops and bidirectional edges.
%In directed uncorrelated random networks with zero clustering coefficient and no bidirectional edges, we have $s_{t}(i) = s_{in}(i) + s_{out}(i)$.

If we neglect the edge directness then the function $C(i \rightarrow j)$ becomes symmetric, $C(i \rightarrow j)=C(j \rightarrow i) = C(i,j)$. Note that in this case $C(i,j)=1$ if
$i$ and $j$ belong to the same cluster $C_{\alpha}$.
%Otherwise, $C(i,j)=0$.
Therefore, according to Eq. (\ref{eq: 3}), the total number $s_{t}(i)$ of vertices reachable from vertex $i$ equals $N(C_{\alpha})-1$ where $N(C_{\alpha})$ is the size the cluster $C_{\alpha}$ to which $i$ belongs.

Below we show that the two-point connectivity function $C(i \rightarrow j)$ is a very useful mathematical object for quantifying the response of a network to the addition and pruning of edge and vertices.

%%%%%%%%%%%%%%%%%%%%%%%%%%%%%%%%%%%%%%%%%%%%%%%%%%%%%%%%%%%%%%%%
\section{Susceptibility of directed networks}
\label{susceptibility}

It is well known that the one-state Potts model is equivalent to the ordinary percolation model \cite{stauffer1994introduction} in undirected networks
\cite{kasteleyn1969phase,wu1982potts}. The Ising model is a particular case of the two-state Potts model. In the Ising model, the susceptibility, $\chi=dM / dH$, quantifies the sensitivity of the magnetization $M$ to an applied magnetic field $H$. The susceptibility is related with the irreducible two-spin correlation function $C(i,j)$ as follows:
\begin{eqnarray}
   \chi &=& \frac{1}{N}\sum_{i,j} C(i,j),
  \label{eq: Ising1} \\
 C(i,j)&=&\langle \sigma_i \sigma_j \rangle_{T} - \langle \sigma_i \rangle _{T} \langle \sigma_j \rangle_{T},
  \label{eq: Ising2}
\end{eqnarray}
where $\langle \sigma_i \rangle _{T}$ stands for the averaging of spin $\sigma_i$ over the Gibbs ensemble (see, for example, in \cite{parisi1988statistical}). The local magnetization is nonzero (i.e., $\langle \sigma_i \rangle _{T} \neq 0$) in the ordered phase at zero magnetic field. The zero-field susceptibility diverges at the critical point signaling a continous phase transition into the ordered phase.

In directed networks there are two connectivity functions, $C(i\rightarrow j)$ and $C(i,j)$,  defined in Sec. \ref{corr function}. Using the equivalence of the percolation model to the one-state Potts model,
%and the two-point connectivity functions,
we introduce two susceptibilities $\chi_{d}$ and $\chi_{t}$,
\begin{eqnarray}
  \chi_{d}&\equiv& \frac{1}{N(F)} \sum_{i,j\in F} C(i\rightarrow j),
  \label{eq: 7} \\
 \chi_t &=& \frac{1}{N(F)} \sum_{i,j\in F} C(i,j),
 \label{eq: 8}
\end{eqnarray}
where $N(F)$ is the number of vertices in the finite components $F$. In these equations
%Eqs. (\ref{eq: susc_d1}) and (\ref{eq: 8})
the sum is only over vertices belonging to  $F$. Vertices belonging to the giant component $G_{IO}$, which is the order parameter for the directed percolation, are not present in the sum similar to the subtraction of the order parameter in the susceptibility $\chi$ of the Ising model. If the giant components are absent then $F=\mathcal{G}$ and  the sum in Eqs. (\ref{eq: 7}) and (\ref{eq: 8}) is over all vertices in the considered network $\mathcal{G}$. In Sec. \ref{network response}, we will show that the susceptibilities $\chi_{d}$ and $\chi_{t}$  quantify the response of directed networks to the addition and pruning of directed and bidirectional edges and vertices. Their divergence signals directed percolation phase transition, i.e., the appearance (or disappearance) of the giant component $G_{S}$.

Using Eqs. (\ref{eq: 3})-(\ref{eq: 5}), we can write $\chi_{d}$ and $\chi_{t}$ in a form
\begin{eqnarray}
  \chi_{d} &= & \frac{1}{N(F)} \sum_{i\in F} [1+ s_{out}^{(F)}(i)] = 1+\langle s_{out}^{(F)}\rangle
  \nonumber \\
  &= & \frac{1}{N(F)} \sum_{i\in F} [1+ s_{in}^{(F)}(i)] = 1+\langle s_{in}^{(F)}\rangle,
  \label{eq: 9} \\
  \chi_t &=& \frac{1}{N(F)} \sum_{i\in F} [1+ s_{t}^{(F)}(i)]= 1+\langle s_{t}^{(F)}\rangle,
  \label{eq: 10}
\end{eqnarray}
where the quantities
\begin{eqnarray}
  s_{in}^{(F)}(i) &=& \sum_{j \in F \backslash i}C(j \rightarrow i),
  \nonumber \\
  s_{out}^{(F)}(i) &=& \sum_{j \in F\backslash i} C(i \rightarrow j),
  \nonumber \\
  s_{t}^{(F)}(i) &=& \sum_{j \in F \backslash i}C(i,j),
  \label{eq: 11}
\end{eqnarray}
are the sizes of the individual in-component, out-component, and the total component, respectively, of vertex $i$ in  $F$.
%%%
Note that
%when calculating the quantities $\langle s_{in}^{(F)}\rangle$, $\langle s_{out}^{(F)}\rangle$, and $\langle s_{t}^{(F)}\rangle$
in Eq. (\ref{eq: 11}) only vertices $j$, which are reachable from $i$ and which belong to $F$, are taken into account. In the general case, the individual in- or out-component of vertex $i  \in F$ can have a nonzero intersection with either $IN$ or $OUT$ as one can see in Fig. \ref{fig: new edge Gin-out}(b). These intersections are excluded from the summation in Eq. (\ref{eq: 11}).
%%%
The quantities $\langle s_{in}^{(F)}\rangle$, $\langle s_{out}^{(F)}\rangle$, and $\langle s_{t}^{(F)}\rangle$ are their mean values,
\begin{eqnarray}
  \langle s_{in(out)}^{(F)} \rangle &\equiv & \frac{1}{N(F)}\sum_{i \in F} s_{in(out)}^{(F)}(i),
  %= \frac{1}{N(F)}\sum_{i,j\in F}C(i \rightarrow j)-1,
  \nonumber \\
  \langle s_{t}^{(F)} \rangle &\equiv & \frac{1}{N(F)}\sum_{i \in F} s_{t}^{(F)}(i).
  %\frac{1}{N(F)}\sum_{i,j\in F}C(i,j)-1.
    \label{eq: 12}
\end{eqnarray}
According to Eqs. (\ref{eq: 9}) and (\ref{eq: 10}),  $\chi_{d}$ equals the mean size of the in-component (or out-component) of a randomly chosen vertex in $F$, also including this vertex but excluding an intersection with $IN$ or $OUT$. $\chi_{t}$ is the mean total size of the in- and out-components of a randomly chosen vertex $F$, including this vertex but excluding an intersection with $IN$ or $OUT$.
Using Eq. (\ref{eq: 2}), we obtain a relation between $\chi_{d}$ and $\chi_{t}$,
\begin{equation}
 \chi_t =2\chi_{d} - \frac{1}{N(F)} \sum_{i,j\in F} C(i \rightarrow j)C(j \rightarrow i).
  \label{eq: 13}
\end{equation}

Equations (\ref{eq: 9}) and (\ref{eq: 10}) show that the susceptibilities $\chi_{d}$ and $\chi_{t}$ are determined by mean sizes of the in- and out-components of vertices. These parameters characterize local properties of vertices, while  Eqs. (\ref{eq: 7}) and (\ref{eq: 8}) relate $\chi_{d}$ and $\chi_{t}$ with the two-point connectivity function, which contains global information about the network connectivity.
%
%Using Eq. (\ref{eq: 2}), we obtain a relation between $\chi_{d}$ and $\chi_{t}$,
%\begin{equation}
% \chi_t =2\chi_{d} - \frac{1}{N(F)} \sum_{i,j\in F} C(i \rightarrow j)C(j \rightarrow i).
%  \label{eq: 13}
%\end{equation}

Let us introduce the susceptibilities
\begin{eqnarray}
 \chi_{in}^{(S)}&\equiv &\frac{1}{N(IN)}\sum_{i,l \in IN}C(i \rightarrow l)
       \label{eq: 25} \\
    &=& 1+ \langle s_{in}^{(IN)} \rangle,
    \nonumber \\
  \chi_{out}^{(S)}&\equiv &\frac{1}{N(OUT)}\sum_{i,l \in OUT}C(l \rightarrow i)
     \label{eq: 26} \\
  &=& 1+ \langle s_{out}^{(OUT)} \rangle,
   \nonumber
    \end{eqnarray}
where $\langle s_{in}^{(IN)} \rangle$ and $\langle s_{out}^{(OUT)} \rangle$ are the mean sizes of the individual in- and out-components of a randomly chosen vertex in the $IN$ and $OUT$, respectively. Note that $\langle s_{in}^{(IN)} \rangle$ and $\langle s_{out}^{(OUT)} \rangle$ are finite in contrast to $\langle s_{out}^{(IN)} \rangle$ and $\langle s_{in}^{(OUT)} \rangle$ that are giant, i.e., of the order of $O(N)$ in the large size limit (see Sec. \ref{corr function}). In Sec. \ref{network response} we show that $\chi_{in}^{(S)}$ and $\chi_{out}^{(S)}$ quantify the response of $IN$, $OUT$, and $G_S$  to the addition or pruning of edges,

We also introduce the probability distribution functions $\Pi_{in}^{(F)}(s)$ and $\Pi_{out}^{(F)}(s)$ of the individual in- and out-components $s_{in}^{(F)}(i)$ and $s_{out}^{(F)}(i)$
%of vertices $i$ in $F$
given by Eq. (\ref{eq: 11}),
\begin{equation}
  \Pi_{in(out)}^{(F)}(s)\equiv \frac{1}{N(F)} \sum_{i \in  F} \delta_{s, s_{in(out)}^{(F)}(i)},
  \label{eq: 14}
\end{equation}
where $\delta_{s,s'}$ is the Kronecker delta.
The normalization condition is $\sum_{s=0}^{\infty} \Pi_{in (out)}^{(F)}(s)=1$. Thus, we can write
\begin{eqnarray}
  \chi_{d}&=&1+\sum_{s=1}^{\infty} s\Pi_{in}^{(F)}(s),
  \nonumber \\
  &=& 1+\sum_{s=1}^{\infty} s\Pi_{out}^{(F)}(s).
  \label{eq: 15}
\end{eqnarray}
This equation shows that the divergence of $\chi_{d}$ is due to the divergence of the first moments of $\Pi_{in}^{(F)}(s)$ and $\Pi_{out}^{(F)}(s)$ at the critical point.
In Sec. \ref{uncorrelated net} we will show that in uncorrelated random directed networks the distribution functions $\Pi_{in}^{(F)}(s)$ and $\Pi_{out}^{(F)}(s)$  have a power law behavior, $\Pi_{in}^{(F)}(s), \Pi_{out}^{(F)}(s) \propto 1/s^{3/2}$, at the critical point.
%This results in the divergence of the first moment of these functions.
In a similar way we introduce the probability distribution function $\Pi_{t}^{(F)}(s)$ of the total size $s_{t}(i)$ of the individual in- and out-components of vertices $i \in F$, excluding an intersection with $IN$ or $OUT$.

In the case of undirected networks, there is only one susceptibility $\chi = \chi_d =\chi_t $.  %susceptibilities $\chi_d$ and $\chi_t$ are equal, $\chi_d =\chi_t \equiv \chi$.
The sum over vertices $i$ and $j$ in Eqs. (\ref{eq: 7}) and (\ref{eq: 8}) can be written as the sum over all finite disconnected clusters $C_\alpha$ in $\mathcal{G}$, except the giant connected component (i.e., except the undirected version of $G_W$),
\begin{equation}
  \chi=\frac{1}{N(C)} \sum_{\alpha} \sum_{i,j \in C_\alpha} C(i,j)=\frac{1}{N(C)} \sum_{\alpha} S_{\alpha}^2.
  \label{eq: 16}
\end{equation}
Here $N (C)$ is the total number of vertices belonging to finite disconnected clusters $C$, i.e., $N(C) = \sum_{\alpha} S_{\alpha}= N(\mathcal{G})-N(G_W)$ where $S_{\alpha} \equiv N(C_{\alpha})$ is the size of a finite cluster $C_{\alpha}$.
According to Eq. (\ref{eq: 16}) the susceptibility $\chi$ of an undirected network is the mean size of a finite cluster to which a randomly chosen vertex belongs. This result is consistent with \cite{stauffer1994introduction}.

\section{Network response to the addition and pruning}
\label{network response}

In this section we consider the response of directed networks having the structure shown in Fig. \ref{fig:structure} to the addition and pruning of edges and vertices.
Actually, we mainly consider the addition of edges. Pruning of edges is the process inverse to the edge addition in the following sense. Structural changes caused by the random pruning of edges are inverse to structural changes caused by the addition of edges at random. The addition and pruning of vertices can be considered in the same way as the addition and pruning of edges. We show that a sensitivity of the different parts of directed networks is quantified by the susceptibilities introduced in Sec. \ref{susceptibility}. Finally, we find analytically behavior of the susceptibilities of uncorrelated random directed networks, when vertices (or edges) are removed at random.
%will be considered in detail in the next section
%in the case of uncorrelated random directed complex networks.

\subsection{Response to edge addition below the percolation point}
\label{response}

First let us consider structural changes caused by the addition of edges to a directed network $\mathcal{G}$ when the network has no giant strongly connected component and there are only finite directed components and finite disconnected clusters, i.e., $F=\mathcal{G}$. In this case, the individual in- and out-components, $s_{in}(i)$ and $s_{out}(i)$, of any vertex $i \in \mathcal{G}$ are finite. The addition of new edges increases
%$s_{in}(i)$ and $s_{out}(i)$
the individual in- and out-components of vertices. This is the process that leads to the appearance of $G_S$ at the critical percolation point.

\begin{figure}[htpb!]
\centering
\includegraphics[width=6cm,angle=0.]{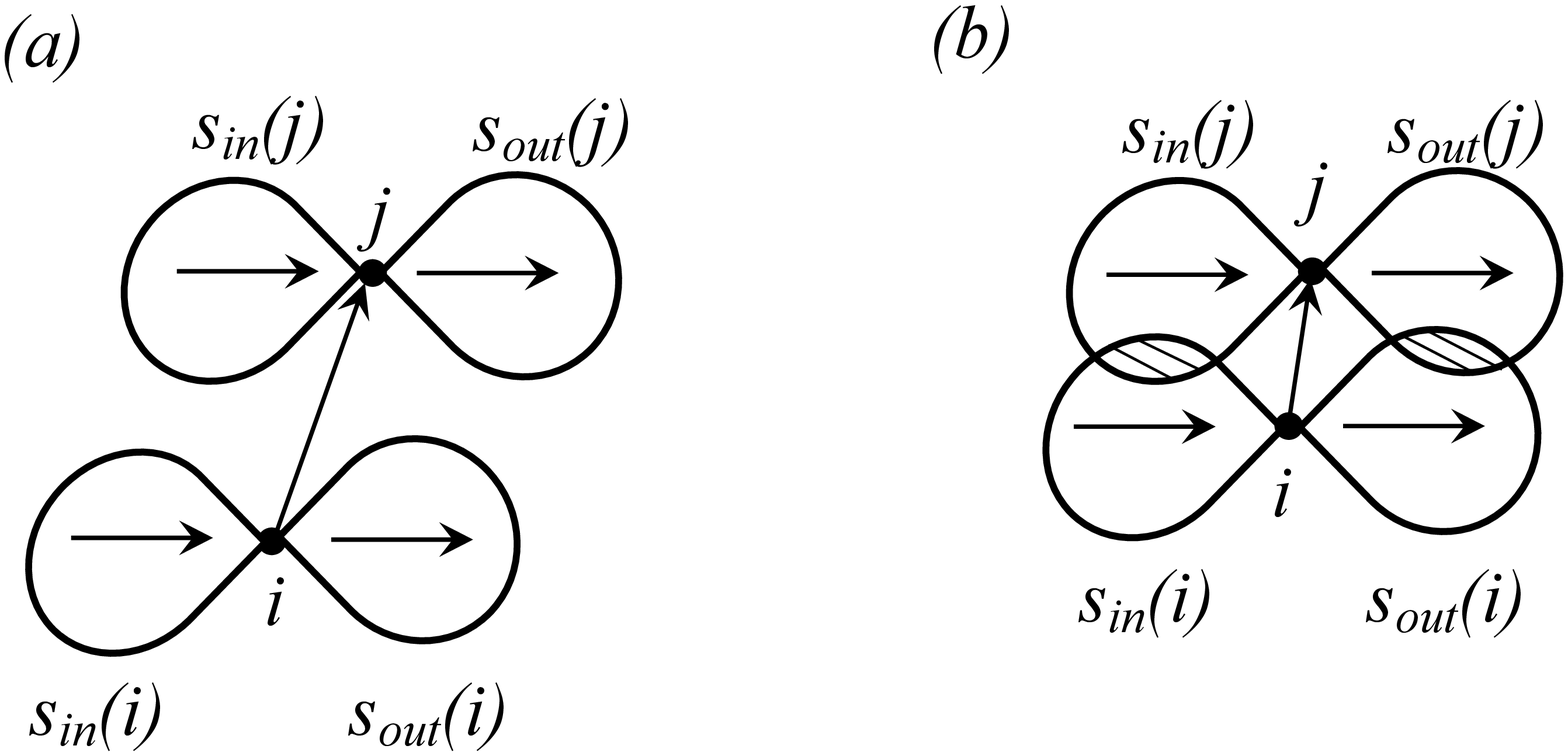}
\caption{The addition of the edge $(i \rightarrow j)$ directed from vertex $i$ to vertex $j$ increases the out-component $s_{out}(i)$ of $i$ by an amount $s_{out}(j)$ and also increases the in-component $s_{in}(j)$ of $j$ by an amount $s_{in}(i)$. (a) In- and out-components of $i$ and $j$ do not intersect each other. (b) There are intersections between the in-and out-components.}
\label{fig: new edge 1}
\end{figure}

Let us choose at random two vertices $i$ and $j$ and add a directed edge $(i \rightarrow j)$ from $i$ to $j$ (see Fig. \ref{fig: new edge 1}). If $j$ does not belong to the out-component of $i$, or $i$ does not belong to the in-component of $j$, then this edge increases the out-component of vertex $i$ and the in-component of vertex $j$ by values
\begin{eqnarray}
\Delta s_{out}(i)&{=}& 1{+}s_{out}(j){-} N(s_{out}(i) \cap  s_{out}(j)),
  \nonumber \\
\Delta s_{in}(j)&{=}& 1{+}s_{in}(i) {-} N(s_{in}(i) \cap  s_{in}(j)),
  \label{eq: 17}
  \end{eqnarray}
respectively.
Here the number of vertices lying in the intersections of the out- and in-components of vertices $i$ and $j$ is subtracted,
\begin{eqnarray}
%1{+}s_{out}(j)&=&\sum_{k} C(j {\rightarrow} k),
%\nonumber \\
%1{+}s_{in}(i)&=&\sum_{k} C(k {\rightarrow} i),
%\nonumber \\
  N(s_{out}(i) \cap  s_{out}(j))&=&\sum_{k} C(j\rightarrow k) C(i \rightarrow k),
  \nonumber \\
  N(s_{in}(i) \cap  s_{in}(j))&=&\sum_{k} C(k \rightarrow i) C(k \rightarrow j).
  \label{eq: 18}
\end{eqnarray}
The mean values of $\Delta s_{out}(i)$ and $\Delta s_{in}(j)$, averaged over pairs of vertices $i$ and $j$, are
\begin{eqnarray}
  \langle \Delta s_{out} \rangle &{=}& \frac{1}{N^2}\sum _{i,j \in \mathcal{G} } \Delta s_{out}(i)
 \nonumber \\
 &{=}& \frac{1}{N^2} \sum_{i,j,k \in \mathcal{G}}  [1 {-} C(i {\rightarrow} j)] C(j {\rightarrow} k)[1 {-} C(i {\rightarrow} k)],
  \nonumber \\
 \langle \Delta s_{in} \rangle &{=}& \frac{1}{N^2}\sum _{i,j \in \mathcal{G} } \Delta s_{in}(j)
  \nonumber \\
   &{=}&\frac{1}{N^2} \!\!\! \sum_{i,j,k \in \mathcal{G}} \!\!\! [1{-} C(i {\rightarrow} j)] C(k{\rightarrow} j)[1 {-} C(k{\rightarrow} i)].
  \label{eq: 18b}
  \end{eqnarray}
The multiplier $1{-} C(i {\rightarrow} j)$ in these equations takes into account the fact that if $j$ belongs to the out-component of $i$ (or equivalently $i$ belongs to the in-component of $j$), then the edge addition gives no contribution to $\Delta s_{out}(i)$ and $\Delta s_{in}(j)$.
%%%%%%%%%%%%%%%%%%%%%%%%%%%%%%%%%%%%%%%%%%%%%%%%
Assuming that all moments of the probability distribution functions $\Pi_{in}(s)$ and $\Pi_{out}(s)$ are finite, we find that the intersections between the in- and out-components give a contribution of order $O(1/N)$ to  $\langle \Delta s_{in} \rangle$ and $\langle \Delta s_{out} \rangle$. Thus, in the infinite size limit $N \rightarrow \infty$, we obtain
  \begin{equation}
  \langle \Delta s_{in} \rangle =\langle \Delta s_{out} \rangle = 1{+}\langle s_{in}\rangle =1{+}\langle s_{out}\rangle= \chi_{d}
  \label{eq: 19}
\end{equation}
This equation shows that the susceptibility $\chi_{d}$ determines an increase of the individual in- and out-components of vertices due to the addition of one edge at random. The larger $\chi_{d}$ the stronger the network response to edge addition.

Let us add a bidirectional edge between a randomly chosen vertices $i$  and $j$. If $i$ does not belong to the individual total component of vertex $j$, and vice versa, then the addition of this edge increases the total size of the in- and out-components of $i$ and $j$ by values
\begin{eqnarray}
\Delta s_{t}(i)&{=}&1{+} s_{t}(j)- \sum_{k}C(j,k)C(i,k),
  \nonumber \\
\Delta s_{t}(j)&{=}&1{+} s_{t}(i)- \sum_{k}C(j,k)C(i,k).
  \label{eq: 20}
  \end{eqnarray}
Here, the intersections of the in- and out-components of $i$ and $j$ are subtracted. Assuming that all moments of the probability distribution function $\Pi_{t}(s)$ are finite in the infinite size limit $N \rightarrow \infty$, we find
\begin{equation}
 \langle \Delta s_{t} \rangle = 1+ \langle s_{t} \rangle =\chi_{t}.
  \label{eq: 22}
  \end{equation}
Therefore,  the susceptibility $\chi_{t}$ quantifies the response to the addition of a bidirectional edge at random.

\subsection{Response to edge addition above the percolation point}
\label{response above}

If a directed network $\mathcal{G}$ has nonempty $G_S$, $IN$, and $OUT$,
then the result of the addition of new edges between two vertices depends on the properties of the network parts to which these vertices belong.
The addition of edges between two vertices in $F$ is described by  Eq. (\ref{eq: 18b}) where we must sum over pairs of vertices $(i,j) \in F$. This process leads to the response Eq. (\ref{eq: 19}) determined by the susceptibility $\chi_d$, Eq. (\ref{eq: 9}). Below we only consider some particular cases of the addition of edges in order to demonstrate that a response of the network is quantified by the susceptibilities. A detailed analysis of the impact of edge pruning on the sizes of $G_S$, $IN$, $OUT$, and $F$ can be performed for a directed uncorrelated random network (see Sec. \ref{uncorrelated net}).

\subsubsection{Impact on $IN$ and $OUT$}

\begin{figure}[htpb!]
\centering
\includegraphics[width=5cm,angle=0.]{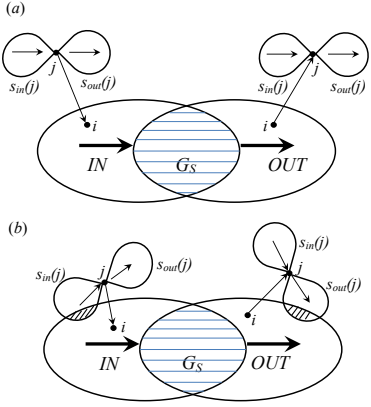}
\caption{ Addition of the edge ($i \rightarrow j$) from vertex $i \in OUT$ (or $G_S$) to vertex $j \in F$ and the edge ($j \rightarrow i$) from vertex $j \in F$ to vertex $i \in IN$ (or $ G_S$). (a)  There are no intersections between  the in- and out-component of vertex $j \in F$ and  $IN$  and $OUT$. (b) The in- and out-components of vertex $j$ intersect  $IN$ and $OUT$ (the shaded regions).}
\label{fig: new edge Gin-out}
\end{figure}

The addition of edges between pairs of vertices $(i,j) \in IN$ or $(i,j) \in OUT$ does not change the sizes of $IN$ and $OUT$.
Let us choose randomly a vertex $i \in G_{out}=OUT \cup G_S$ and a vertex $j$ in the finite component $F$ (tendrils, tubes, and finite disconnected clusters).
We add an edge $(i \rightarrow j)$ directed from $i$  to $j$ (see Fig. \ref{fig: new edge Gin-out}).
As a result,  all vertices in the out-component $s_{out}(j)$ of $j$ become a part of $OUT$ because now there is a directed path from vertices in $G_S$ through $i$ to any vertex in $s_{out}(j)$.
%Due to the addition of the edge $i \rightarrow j$ from vertex $i \in G_{out}=OUT \cup G_S$) to vertex $j \in F$ all vertices in the out-component $s_{out}(j)$ of $j$ become a part of the $OUT$ because now there is a directed path from $G_S$ through $i$ to these vertices.
Vertices belonging the intersections between
%$s_{in}(j)$ and $IN$,
$s_{out}(j)$ and $OUT$ (the shaded regions in Fig. \ref{fig: new edge Gin-out} (b)) already belong to $OUT$ and should not be considered.
Thus, $OUT$ increases by the value,
\begin{equation}
  \Delta N(OUT) = \frac{1}{N(F)} \sum_{j,k \in F} C(j \rightarrow k)= \chi_d,
  \label{eq: 23}
\end{equation}
where we used the definition Eq. (\ref{eq: 7}). Note that the edge $(j \rightarrow i)$ does not change the size of $OUT$.

In the same way, we find the response of $IN$ to the addition of an edge $(j \rightarrow i)$ directed from vertex $j \in F$ to vertex $i \in IN \cup G_S$ chosen at random. In this case
\begin{equation}
  \Delta N(IN) =\frac{1}{N(F)} \sum_{j,k \in F} C(k \rightarrow j)= \chi_d.
  \label{eq: 23IN}
\end{equation}
Therefore the susceptibility $\chi_{d}$ quantifies the response of $IN$  and $OUT$ to the addition of edges between vertices in $IN$  and $OUT$ and vertices in $F$.

\subsubsection{Impact on $G_S$ }

\begin{figure}[htpb!]
\centering
\includegraphics[width=4cm,angle=0.]{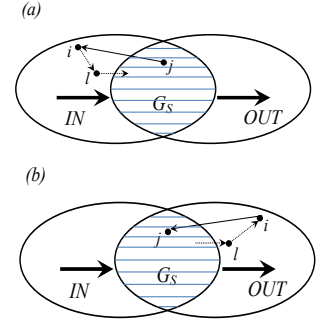}
\caption{(Color online) (a) The addition of an edge directed from vertex $j \in G_{S}$ to vertex $i \in IN$ increases the size of the giant strongly connected component $G_{S}$ due to  vertices $l \in IN$ reachable from $i$. (b) The addition of an edge directed from vertex $i \in OUT$ to vertex $j \in G_{S}$ increases $G_{S}$ due to vertices $l \in OUT$ from which $i$ can be reached.}
\label{fig: new edge GS-GIN-GOUT}
\end{figure}

The addition of edges between pairs of vertices $i$  and $j$ belonging to the giant strongly connected component $G_S$ does not change its size.
Let us choose at random two  vertices, one vertex $i$ in $IN$ ($i \in IN$) and the other vertex $j$ in $G_S$ ($j \in G_{S}$), and make a directed link ($j \rightarrow i$) from $j$ to $i$ (see Fig. \ref{fig: new edge GS-GIN-GOUT}). One can see that all vertices $l \in IN$ for which $C(i\rightarrow l) =1$ become a part of $G_{S}$ because they satisfy the criterion formulated in Sec. \ref{structure}, i.e., these vertices can now reach any vertex in $G_S$ by following edges either backwards or forwards [see Fig. \ref{fig: new edge GS-GIN-GOUT}(a)]. Thus the size $N(G_S)$ of $G_S$ increases by a value $\sum_{l \in IN} C(i \rightarrow l)$. On average,
this value per one added edge is
\begin{equation}
\Delta N(G_S) = \frac{1}{N(IN)}\sum_{i,l \in IN}C(i \rightarrow l)=\chi_{in}^{(S)},
  \label{eq: 24}
\end{equation}
where we used Eq. (\ref{eq: 25}). Note that the addition of a directed  edge ($i \rightarrow j$)  does not change $G_S$.

\begin{figure}[htpb!]
\centering
\includegraphics[width=3.5cm,angle=0.]{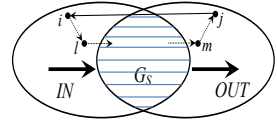}
\caption{(Color online) The addition of an edge directed from vertex $j \in OUT$ to vertex $i \in IN$ increases the size of the giant strongly connected component $G_{S}$ at the expense of  vertices $l \in IN$ reachable from $i$ and vertices $m \in OUT$ which can reach $j$ by following edge directions.}
\label{fig: new edge for S}
\end{figure}

Let us add a directed edge ($i \rightarrow j$) from vertex $i \in OUT$ to vertex $j \in G_{S}$. Then all vertices $l \in OUT$ for which $C(l\rightarrow i) =1$ will belong to $G_{S}$ [see Fig. \ref{fig: new edge GS-GIN-GOUT}(b)]. In this case, the size $N(G_S)$ of $G_S$ increases by a value
\begin{equation}
\Delta N(G_S) = \frac{1}{N(OUT)}\sum_{l \in OUT} C(l \rightarrow i)=\chi_{out}^{(S)},
  \label{eq: 24IN}
\end{equation}
where we used Eq. (\ref{eq: 26}). The addition of a directed  edge ($j \rightarrow i$)  does not change the size of $G_S$.
One can also increase $G_S$ by adding an edge $(j\rightarrow i)$ directed from vertex  $j \in OUT$ to vertex $i \in IN$ (see Fig. \ref{fig: new edge for S}). On average, the addition of this edge  increases the size of $G_S$ by the value $\chi_{in}^{(S)}+\chi_{out}^{(S)}$ per one added edge while the addition of the edge $(i\rightarrow j)$  does not change $G_S$. This edge is an edge-tube. Note that the addition of edges represented in Figs. \ref{fig: new edge GS-GIN-GOUT} and \ref{fig: new edge for S} results in the formation new feedback loops in the modified $G_S$.

One can increase the size of the giant strongly connected component $G_S$ by choosing at random a vertex $j$ belonging to $F$ and connecting it by a bidirectional edge with any vertex $i \in G_S$. $G_S$ increases on average by the value
\begin{equation}
\Delta N(G_S) = \frac{1}{N(F)}\sum_{j \in F}[1+s^{(F)}_{t}(j)] = \chi_t.
  \label{eq: 24t-G}
\end{equation}

The susceptibilities $\chi_{in}^{(S)}$ and $\chi_{out}^{(S)}$ quantify the sensitivity of the giant strongly connected component $G_S$ to the addition of one edge at random. According to Eqs. (\ref{eq: 25}) and (\ref{eq: 26}), these susceptibilities are determined by the statistics of the individual finite in- and out- components of vertices in $IN$ and $OUT$, respectively. The susceptibilities are related with the probability distribution functions  $\Pi_{in}^{(G)}(s)$  and $\Pi_{out}^{(G)}(s)$ of $s_{in} (IN)$ and $s_{out} (OUT)$ in $IN$ and $OUT$, respectively, similarly to Eq. (\ref{eq: 15}).
The susceptibilities $\chi_{in}^{(S)}$ and $\chi_{out}^{(S)}$ have no analogy in undirected networks. They exist only in the phase with the giant strongly connected component $G_S$. Below we will show that $\chi_{in}^{(S)}$ and $\chi_{out}^{(S)}$ diverge at the critical percolation point. This divergence is due to the divergence of $\langle s_{in}^{(IN)} \rangle$ and $\langle s_{out}^{(OUT)} \rangle$.

\subsubsection{Addition of a directed edge at random}
\label{uncorrelated net}

Let us add a directed edge between two randomly chosen vertices $i$ and $j$ in $\mathcal{G}$ and find how it changes sizes of $F$, $IN$, $OUT$, and $G_S$. The edge can be directed with the probability $1/2$ either from $i$ to $j$ or from $j$ and $i$. The processes in Figs. \ref{fig: new edge Gin-out}-\ref{fig: new edge for S} give
\begin{eqnarray}
\Delta N(F) &=& - \frac{1}{2} \chi_{d} S_{F} (S_{S}+S_{IN}) - \frac{1}{2} \chi_{d} S_{F} (S_{S}+S_{OUT}),
 \nonumber \\
 \Delta N(IN) &=& \frac{1}{2} \chi_{d} S_{F} (S_{S}+S_{IN}) - \frac{1}{2} \chi_{IN}^{(S)} S_{S} S_{IN}
 \nonumber \\
  &-& \frac{1}{2}[\chi_{in}^{(S)} + \chi_{out}^{(S)}] S_{OUT} S_{IN},
 \nonumber \\
  \Delta N(OUT) &=& \frac{1}{2} \chi_{d} S_{F} (S_{S}+S_{OUT}) - \frac{1}{2} \chi_{OUT}^{(S)} S_{S} S_{OUT}
 \nonumber \\
  &-& \frac{1}{2}[\chi_{in}^{(S)} + \chi_{out}^{(S)}] S_{OUT} S_{IN},
 \nonumber \\
\Delta N(G_S) &=& \frac{1}{2} \chi_{in}^{(S)} S_{IN} S_{S}+ \frac{1}{2} \chi_{out}^{(S)} S_{OUT} S_{S}
 \nonumber \\
&+&\frac{1}{2}[\chi_{in}^{(S)} + \chi_{out}^{(S)}] S_{OUT} S_{IN}.
  \label{eq: 24-6}
\end{eqnarray}
where $S_{F} \equiv N(F)/ N$, $S_{IN} \equiv N(IN)/ N$, $S_{OUT} \equiv N(OUT)/ N$, and $S_{S} \equiv N(G_S)/ N$ are the fraction of vertices belonging to $F$, $IN$, $OUT$, and $G_S$.
Therefore, at $p > p_c$ the addition of an edge at random decreases the size of $F$ due to the processes in Fig. \ref{fig: new edge Gin-out} while the size of $G_S$ increases due to the processes in Figs. \ref{fig: new edge GS-GIN-GOUT} and \ref{fig: new edge for S}. The sizes of  $IN$ and $OUT$ can both increase and decrease  in dependence on the values of the negative contribution from the processes in Figs.  \ref{fig: new edge GS-GIN-GOUT} and \ref{fig: new edge for S} and the positive contribution from the processes in Fig. \ref{fig: new edge Gin-out}. One can see this behavior in Figs. \ref{fig:real} and \ref{fig:real2} displaying results of our simulations for some real and synthetic directed complex networks.

\section{Susceptibility of randomly damaged uncorrelated directed networks}
\label{uncorrelated net}

In the previous section we introduced the susceptibility quantifying the sensitivity of different parts of directed networks to damage. In this section we find analytically the susceptibility of randomly damaged uncorrelated random directed complex networks. In this kind of complex networks, degree-degree correlations between different vertices are absent.
%and the clustering coefficient is zero in the infinite size limit.
Moreover, such complex networks have locally tree-like structure.

Let us consider the case of a randomly damaged uncorrelated directed network $\mathcal{G}$ and  $p$ is the occupation probability of vertices in this network. In other wards, vertices are removed with the probability $1-p$ and remain in the network with the probability $p$.
%This kind of random damage impacts the structure of the directed network under consideration.  In particular,
With increasing the fraction of removed vertices (this corresponds to decreasing the occupation probability $p$) the giant strongly connected component $G_{S}$ decreases while the finite directed components $F$ grow. The network undergoes the percolation phase transition at the critical point $p_c$ at which $G_S$ disappears. At $p < p_c$ there are only finite directed components.
Structural changes caused by random removal of vertices or edges in the directed network  can be described analytically by use of the generating function method \cite{newman2001random,dorogovtsev2001giant,schwartz2002percolation,boguna2005generalized}, which
%in the case when the directed network is uncorrelated random complex network.
%For this kind of networks the generating function method
gives exact results for uncorrelated random complex networks in the infinite size limit. Note that the same generating function method can be used for analyzing edge pruning. Real directed networks are correlated and have a finite size and a large clustering coefficient \cite{fagiolo2007clustering,bianconi2008local}. Nevertheless, we expect that even in this case one can use the results obtained by the generating function methods. These results provide a qualitatively correct though approximate description of structural changes caused by damage.

\subsection{Susceptibility $\chi_d$}
\label{chi-d in uncorrelated net}

Let us find the susceptibilities $\chi_{d}$ and  $\chi_{t}$ [see Eqs. (\ref{eq: 9}), (\ref{eq: 10})] of randomly damaged uncorrelated directed networks. On-site correlations between in- and out-degrees are characterized by a function $P(q_i,q_o)$, which is the probability that a randomly chosen vertex has in-degree $q_i$ and out-degree $q_o$. The mean in- and out-degrees are $\langle q_i \rangle \equiv \sum_{q_i, q_o} q_i P(q_i,q_o)$ and $\langle q_o \rangle \equiv \sum_{q_i, q_o} q_o P(q_i,q_o)$, respectively. Note that $\langle q_i \rangle = \langle q_o \rangle$ in any directed network. First we consider the case when all edges are directed and there are no bidirectional edges. The case when there are both directed and bidirectional edges will be considered in Sec. \ref{susc bidirectional}.
Due to the tree-like structure, the total size $s_{t}^{(F)}(i)$ of the in- and out-components of vertex $i \in F$ in Eq. (\ref{eq: 5}) is the sum $s_{in}^{(F)}(i) + s_{out}^{(F)}(i)$ because $s_{in}^{(F)}(i)$ and $s_{out}^{(F)}(i)$ do not intersect each other. Equation (\ref{eq: 13}) gives
\begin{equation}
\chi_{t}=2 \chi_{d}-1.
  \label{eq: 27}
\end{equation}
According to Eq. (\ref{eq: 9}), the susceptibility $\chi_{d}$ is determined by the mean size of the individual in- or out-components ($\langle s_{in}^{(F)} \rangle$ or  $\langle s_{out}^{(F)} \rangle$) of vertices belonging to $F$. In order to find these parameters
we use the generating function method
described in Appendix \ref{generating functions}.
Using Eqs. (\ref{eq: A14}), (\ref{eq: A16 below}), (\ref{eq: 31}), and (\ref{eq: 32}),
we find explicitly the susceptibility $\chi_d$ in uncorrelated random directed complex networks at $p \leq p_c$,
\begin{equation}
\chi_{d} =
%1+ \langle s_{out} (F) \rangle =
1+ \frac{\langle q_o \rangle p p_{c}}{p_c-p},
  \label{eq: 34}
\end{equation}
where $p_c= \langle q_o \rangle/\langle q_i q_o \rangle$, see Eq. (\ref{eq: A15}). Using Eqs. (\ref{eq: A9}), (\ref{eq: A10}), (\ref{eq: A20}), and (\ref{eq: 31}) in Appendices \ref{generating functions} and \ref{generating functions: finite components}, we find $\langle s_{out}^{(F)} \rangle$ and the critical behavior of susceptibility $\chi_{d} $  above the critical point ($p > p_c$),
\begin{equation}
 \chi_{d} \approx
 %\langle s_{out} (F) \rangle \approx
 \frac{\langle q_o \rangle p p_{c}}{3(p-p_c)}.
  \label{eq: 35}
\end{equation}
Thus, according to Eqs. (\ref{eq: 34}) and (\ref{eq: 35}), the susceptibility $\chi_{d}$
%and the mean size $\langle s_{in (out)} (F) \rangle$ of the in- and out components of vertices in the tendrils and tubes
diverges as $A_\chi/| p-p_c |$ when the directed network approaches the critical percolation point $p_c$ both from below and above. The difference is only in the amplitude $A_\chi$, which is three times smaller at $p >p_c$ in comparison with the one at  $p \leq p_c$, i.e.,
\begin{equation}
 %\frac{\chi_{d}\Big|_{p \rightarrow p_c -0}}{\chi_{d}\Big|_{p \rightarrow p_c +0}} = 3.
 \frac{\chi_{d}(p \rightarrow p_c -0)}{\chi_{d}(p \rightarrow p_c +0)} = 3
  \label{eq: 34/35}
\end{equation}
Our numerical simulations in Sec. \ref{simulations}
confirm the analytical results. Note that in the framework of the phenomenological Landau theory of continuous phase transitions (see, for example, \cite{goltsev2003critical}), the ratio $\chi(p \rightarrow p_c -0)/\chi(p \rightarrow p_c +0)$ is 1 for the percolation transition in contrast to 3 in Eq. (\ref{eq: 34/35}).
%This result shows that some critical properties of percolation transition in directed networks are different from the properties of the ordinary percolation transition in undirected networks.
For comparison, $\chi(T \rightarrow T_c -0)/\chi(T \rightarrow T_c +0)=2$ for the ferromagnetic transition within the mean-field theory.

Based on Eqs. (\ref{eq: 34}), (\ref{eq: 35}), and (\ref{eq: A12}), we conclude that
in  uncorrelated random  directed networks the susceptibility $\chi_d$ and the order parameter $S_{IO}$  demonstrate the critical behavior: $\chi_d \propto |p-p_c|^{-\gamma}$ and $S_{IO} \propto (p-p_c)^{\beta} $ with the standard critical exponents $\gamma =1$ and $\beta=1$. We find the same critical behavior in the uncorrelated random directed  networks with bidirectional edges (see Appendix \ref{susc bidirectional}).

\subsection{Susceptibilities $\chi_{in}^{(S)}$ and $\chi_{out}^{(S)}$ }
\label{susceptibilities for Gs }

Let us find the susceptibilities $\chi_{in}^{(S)}$ and $\chi_{out}^{(S)}$ quantifying the sensitivity of $G_S$ to the addition of edges in directed random uncorrelated networks. According to Eqs. (\ref{eq: 25}) and (\ref{eq: 26}), $\chi_{in}^{(S)}$ and $\chi_{out}^{(S)}$ are determined by the mean size of the individual in- and out-components, $\langle s_{in}^{(IN)} \rangle$ and $\langle s_{out}^{(OUT)} \rangle$ of vertices in $IN$ and $OUT$, respectively.
Using Eqs. (\ref{eq: A9}), (\ref{eq: A10}),  and (\ref{eq: A16}), in the leading order of $(p-p_c)/p_c$ at $p > p_c$,
we find the critical behavior
\begin{equation}
 \chi_{in}^{(S)} {\approx} \chi_{out}^{(S)} {\approx}
 %\langle s_{out} (OUT) \rangle {\approx} \langle s_{in} (G_{in}) \rangle {\approx}
 \frac{\langle q_i q_o \rangle}{\langle q_i \rangle} \frac{ p_{c}^2}{p-p_c},
  \label{eq: 39}
\end{equation}
(see Eqs. (\ref{eq: 36}) and (\ref{eq: 37})).
Using Eqs. (\ref{eq: 35}) and (\ref{eq: 39}), at $p$ near $p_c$  we find the ratio
\begin{equation}
 \frac{\chi_{in}^{(S)}}{\chi_{d}} = \frac{3 \langle q_i q_o \rangle }{\langle q_i \rangle\langle q_o \rangle}.
  \label{eq: 39r}
\end{equation}
In Sec. \ref{simulations} we analyze numerically the critical behavior of  $\chi_{in} ^{(S)}$ and $\chi_{out} ^{(S)}$ in real and synthetic directed networks.

\subsection{Statistics of the in- and out-components of vertices}
\label{statistics in F }

In the case of uncorrelated random complex networks, the distribution functions $\Pi_{in}^{(F)}(s)$ and $\Pi_{out}^{(F)}(s)$ [see Eq. (\ref{eq: 14})]
%of the finite in- and out-components of vertices in the finite components $F$
are related with the generating functions Eqs. (\ref{eq: A17}) and (\ref{eq: A18}) in Appendix \ref{generating functions: finite components}.
Using the analytical method developed in \cite{newman2001random}, we find that these distribution functions have the following asymptotic behavior:
\begin{equation}
 \Pi_{out}^{(F)}(s) \propto Y_{in}^{(F)}(s)
  \propto  \frac{1}{s^{3/2}} e^{-s/s^{*}}
  \label{eq: 40}
\end{equation}
both below and above $p_c$. The parameter $s^{*}$ behaves as
$s^{*} \propto (p-p_c)^{-2}$. At $p$ below $p_c$ the asymptotic behavior Eq. (\ref{eq: 40}) was found in \cite{schwartz2002percolation}. The distribution functions $\Pi_{in}^{(F)}(s)$ and $\Pi_{out}^{(F)}(s)$ have the power-law behavior $s^{-3/2}$ at the critical point $p_c$.
We find the same asymptotic behavior Eq. (\ref{eq: 40}) for the probability distribution functions $\Pi_{in}^{(IN)}(s)$ and $\Pi_{out}^{(OUT)}(s)$ of the in-component $s_{in}^{(IN)}(i)$ and  the out-component $s_{out}^{(OUT)}(i)$ of vertices $i \in OUT$ and $i \in OUT$, respectively.

\section{Susceptibility of real and synthetic directed networks}
\label{simulations}

\begin{figure}[htpb!]
\centering
%\includegraphics[height=5.32cm, angle=0]{sus_curves.eps} \\
%\hspace*{-0.3cm}
%\includegraphics[height=5.2cm, angle=0]{r_curves.eps}
%\hspace*{-0.3cm}
%\includegraphics[height=5.2cm, angle=0]{ratio.eps}
\includegraphics[height=10cm, angle=0]{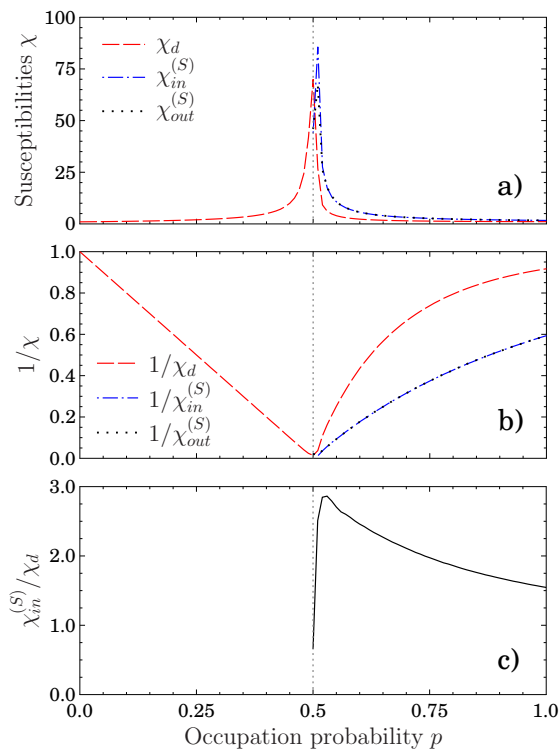} \\
%\ec
%\vspace*{-0.5cm}
\caption[]{\small\label{fig:rasterplotexamp}
(Color online).
(a) Susceptibilities $\chi_d$, $\chi_{in}^{(S)}$, and $\chi_{out}^{(S)}$ versus the occupation probability $p$ in an Erd\H os-R\'enyi network of size $N=10^6$ and the mean in- and out degrees $\langle q_i \rangle =\langle q_o \rangle= 2$.
(b) Reciprocal susceptibilities: $1 / \chi_d$, $1 / \chi_{in}^{(S)}$, and $1 / \chi_{out}^{(S)}$ versus $p$.
(c) The ratio $\chi_{in}^{(S)} / \chi_d $ above $p_c$.
%Results of averaging
%More than 100 network realizations are used for each value of $p$.
100 network realizations were used for each value of p, and in each realization $10\%$ of the vertices were sampled randomly.
}
\label{fig: susc ER graph}
\end{figure}

In this section we discuss the results of our numerical simulations of the susceptibilities $\chi_d$, $\chi^{(S)}_{in}$ , and $\chi^{(S)}_{out}$ in  randomly damaged real and synthetic networks.
%We compare numerical results and our analytical predictions for Erd\H os-R\'enyi networks.
In Sec. IV we showed that these susceptibilities are  determined by the two-point connectivity function $C(i \rightarrow j)$ [see Eqs. (\ref{eq: 7}), (\ref{eq: 25}) and (\ref{eq: 26})]. Unfortunately, it is computationally inefficient to explicitly find $C(i \rightarrow j)$. An alternative numerical method for finding these susceptibilities is to use the fact that, according to Eqs. (\ref{eq: 9}),  (\ref{eq: 25}) and (\ref{eq: 26}), the susceptibilities are determined by the mean size of the finite individual in- and out- components of vertices in the network parts $F$, $IN$, and $OUT$ (see Fig. \ref{fig:structure}). Thus, the statistical analysis of the individual in- and out-components of randomly chosen vertices allows us to find numerically the susceptibilities both above and below the percolation transition. In the simulations, the networks under consideration were randomly damaged, i.e., edges were removed with a probability $1{-}p$ and were retained with an occupation probability $p$. We found $F$, $IN$, $OUT$, and $G_S$ of the damaged networks. Note that since the networks studied in the simulations have a finite size, we considered the largest strongly connected component as $G_S$. Then  we chose a sample subset of vertices uniformly at random from $F$, $IN$, and $OUT$, and determined the sizes of the individual in- and out components of vertices in these components [see Eqs. (\ref{eq: 3})-(\ref{eq: 6}), (\ref{eq: 25}) and (\ref{eq: 26})].
These sizes were averaged over the vertices in the chosen subset, and over many realizations of the damage, to arrive at estimates for the susceptibilities [see Eqs. (\ref{eq: 9}),  (\ref{eq: 25}) and (\ref{eq: 26})].
Figure \ref{fig: susc ER graph} represents results of our simulations for the Erd\H os-R\'enyi networks with uncorrelated in- and out-degrees. The susceptibilities $\chi_d$, $\chi^{(S)}_{in}$, and $\chi^{(S)}_{out}$ demonstrate a sharp maximum that signals the percolation transition.
%and the appearance of the $G_S$ above $p_c$.
Equation (\ref{eq: A15}) predicts that the critical point of the Erd\H os-R\'enyi network is $p_c = \langle q_i \rangle / \langle q_i q_o \rangle = 1 / \langle q_i \rangle$. In the simulations we studied networks with the mean in- and out degrees $\langle q_i \rangle =\langle q_o \rangle= 2$, i.e., $p_c = 0.5$. In this case, equation (\ref{eq: 34}) gives $1 / \chi_d =(p_c - p) / p_c = 1-2p$ in the region $0 \leq p \leq p_c$ where there are only finite directed components and disconnected clusters.  This theoretical prediction is in complete agreement with our simulations in Fig. \ref{fig: susc ER graph}(b) (see the red dashed line at $0 \leq p \leq 0.5$).
The critical behavior of $\chi^{(S)}_{in}$ and $\chi^{(S)}_{out}$ is shown in Fig. \ref{fig: susc ER graph}(a) and \ref{fig: susc ER graph}(b).
%It agrees completely with Eq. (\ref{eq: 39}).
Studying networks of different size $N$, we observed that the maxima of $\chi_d$, $\chi^{(S)}_{in}$, and $\chi^{(S)}_{out}$ increase with increasing $N$ showing the tendency for the divergence in the limit $N \rightarrow \infty$. Results in Fig. \ref{fig: susc ER graph}(b) also show that the susceptibility $\chi_d$ has different slopes at $p$ above and below $p_c$, in agreement with Eqs. (\ref{eq: 35}) and (\ref{eq: 34/35}), which predict $p_c / [(p_c - p) \chi_d] = 1$ at $p \leq p_c$ and $p_c / [(p - p_c) \chi_d] = 3$ at $p > p_c$. At $p > p_c$ the reciprocal susceptibilities $\chi^{(S)}_{in}$ and $\chi^{(S)}_{out}$ behave as follows: $1 / \chi^{(S)}_{in} = 1 / \chi^{(S)}_{out} \approx (p - p_c) / p_c = 2p-1$ in agreement with  Eq. (\ref{eq: 39}) [see the blue dot-dashed line in Fig. \ref{fig: susc ER graph}(b)].  Figure \ref{fig: susc ER graph}(c) displays the ratio $\chi^{(S)}_{in} / \chi_{d}$. One can see that this ratio tends to 3 at $p \rightarrow p_c$ in agreement with Eq. (\ref{eq: 39r}).  $\chi^{(S)}_{in}$ achieves a maximum at $p_{max}$ slightly above $p_c$. This shift of $p_{max}$ from $p_c$  is due to a finite-size effect. It becomes smaller and smaller with increasing $N$.

\begin{figure}[htpb!]
\centering
\includegraphics[width=8cm,angle=0.]{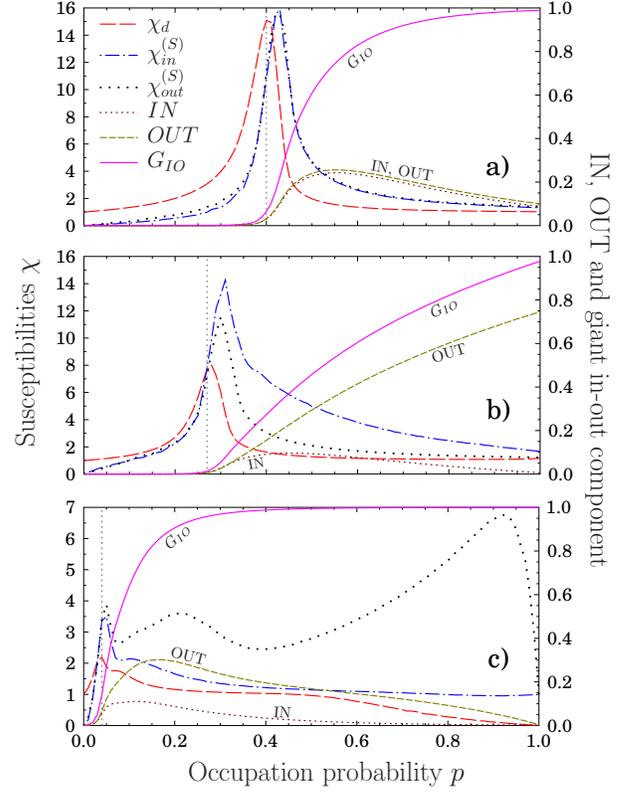}
\caption{(Color online) Susceptibilities $\chi_d$, $\chi^{(S)}_{in}$, and $\chi^{(S)}_{out}$, and the relative sizes of $IN$, $OUT$ and $G_{IO}$ versus the occupation probability $p$: (a) an Erd\H os-R\'enyi network of size $N=10^4$; (b) the Gnutella p2p file sharing network ($N = 62 586, \langle q_{tot} \rangle = 4.726$ \cite{konect:2016:p2p-Gnutella31,konect:ripeanu02}); (c) the neural network of \emph{C. elegans} ($N = 495, \langle q_{tot} \rangle = 32.073$ \cite{jarrell2012connectome}). The vertical dashed lines correspond to $p_c$ determined by the message passing method of \cite{timar2017mapping}.
}
\label{fig:real}
\end{figure}

%Results of our simulations showing
We found  a similar critical behavior of the susceptibilities
%in our simulations for
in the Gnutella p2p filesharing network \cite{konect:2016:p2p-Gnutella31,konect:ripeanu02} and the neural network of C. elegans \cite{jarrell2012connectome} (see Fig. \ref{fig:real} where an Erd\H os-R\'enyi network is also displayed for comparison).
Our analysis of data \cite{jarrell2012connectome} showed that that the main body of the male C. elegans consists of 495 vertices wired by both chemical and electrical synapses. There are 492 nodes in the $G_S$, 1 node in $IN$ and 2 nodes in $OUT$. 
%%%%%%%%%%%%%%%5
%These 3 neurons are the following: neuron with label "AINL" in $IN$. This neuron is in the "Amphid interneurons" and has one outgoing link to "SAAVL" in the "Sublateral cord motor neurons". It has no incoming links. Neuron with label "RIPL" in OUT. This neuron is in "Ring interneurons" and has incoming links from "Sensory neurons", "Amphid interneurons" and "Ring motor neurons". All together it has 8 incoming links and no outgoing links. Neuron with label "gonad" in $OUT$. This neuron is the separate "gonad". It has incoming links from "Ventral cord interneurons", "Male sensory neurons" and "Male interneurons". All together it has 14 incoming links and no outgoing links.
%\cite{[See Supplemental Material for the complete decomposition of a simple directed graph at ]timarSM}
%%%%%%%%%%
In Fig. \ref{fig:real} we show the behavior of the relative sizes of $IN$, $OUT$, and the order parameter $G_{IO}$, which is the union $G_{IO}= G_S \cup IN \cup OUT $ [Eq. (\ref{eq: 1op})], as functions of the occupation probability $p$. The maximum of $\chi_d$ signals the
%appearance of $G_S$, i.e., $G_{IO}$
percolation transition. The position of the maximum agrees very well with $p_c$ predicted by the message passing algorithm of [19].
Notice that the maxima of the susceptibilities $\chi^{(S)}_{in}$ and $\chi^{(S)}_{out}$ are slightly shifted, compared to the maximum of $\chi_d$. This shift is due to a finite size effect similar to the one in the Erd\H os-R\'enyi network in Fig. \ref{fig: susc ER graph} (a).
%In Fig. \ref{fig:real} we also plot the relative sizes of $IN$, $OUT$ and $G_{IO}$ in the damaged networks versus $p$.
In the Erd\H os-R\'enyi network, the sizes of $IN$ and $OUT$ achieve a maximum and then decrease with increasing $p$, in contrast to the strictly monotonic increase of $G_{IO}$. This feature is shared by the \emph{C. elegans} network. However the size of $OUT$ in the Gnutella network increases monotonically, without producing a maximum. The susceptibilities $\chi_d$, $\chi^{(S)}_{in}$, and $\chi^{(S)}_{out}$ of the Gnutella network, after their peak at $p_c$ decrease monotonically, similarly to their behavior in the Erd\H os-R\'enyi network. A striking exception from this rule is the susceptibility $\chi^{(S)}_{out}$ in the \emph{C. elegans} network, which exhibits a strongly non-monotonic behavior [two additional broad maxima above $p_c$, apart from the sharp maximum at $p_c \approx 0.04$ in Fig. \ref{fig:real} (c), dotted line]. This unusual behavior is due to the structural peculiarities of the \emph{C. elegans} network. We suggest that the maximum at $p\approx 0.92$ is due to the chains of bodywall muscle neurons, which all have exactly $2$ outgoing connections (to their neighbors on either side), and multiple in-coming ones. Pruning of some connections between these neurons in the chain increases significantly the $OUT$ and, in turn, increases $\chi^{(S)}_{out}$ even at small damage. The origin of the maximum at the intermediate $p$ ($p \approx 0.2$), is unclear and needs a more detailed analysis of the impact of pruning on the \emph{C. elegans} network.

\begin{figure}[htpb!]
\centering
\includegraphics[height=8cm, angle=0]{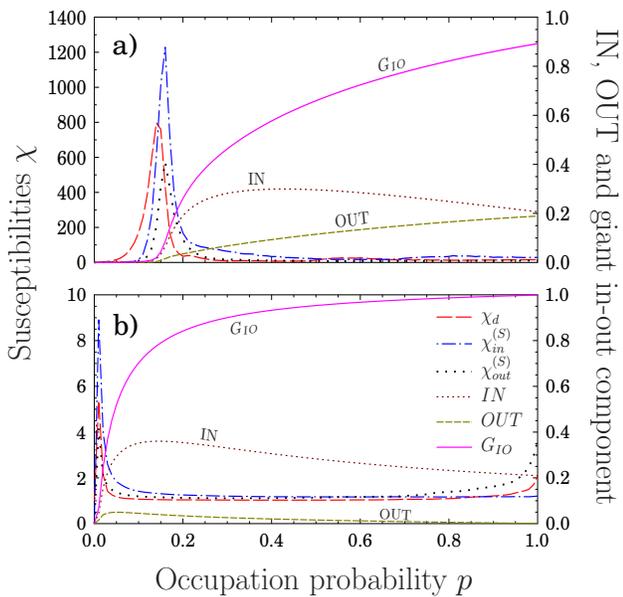} \\
%\hspace*{-0.3cm}
%\includegraphics[height=5.2cm, angle=0]{r_curves.eps}
%\hspace*{-0.3cm}
%\includegraphics[height=5.2cm, angle=0]{ratio.eps}
%\includegraphics[height=4cm, angle=0]{WWW.eps} \\
%\includegraphics[height=4cm, angle=0]{Twitter.eps} \\
%\ec
%\vspace*{-0.5cm}
\caption[]{(Color online).
Susceptibilities $\chi_d$, $\chi^{(S)}_{in}$, and $\chi^{(S)}_{out}$, and the relative sizes of  $IN$, $OUT$ and $G_{IO}$ versus the occupation probability $p$: (a) a sample of the World Wide Web ($N = 875713, \langle q_{in} \rangle = 5.83$ \cite{konect:2016web-Google, konect:leskovec08}), (b) a sample of Twitter ($N = 456631, \langle q_{in} \rangle = 32.53$ \cite{kwak2010twitter}).
}
\label{fig:real2}
\end{figure}

In Figs. \ref{fig:real2} (a) and (b) we present results of our simulations for samples of two well-known, inherently directed networks, the World Wide Web \cite{konect:2016web-Google, konect:leskovec08}  and Twitter \cite{kwak2010twitter}.
Note that in the WWW a directed link from $i$ to $j$ means that there is a hyperlink from site $i$ pointing to site $j$. In Twitter, a directed link from $i$ to $j$ means that $i$ follows $j$.
Both networks have similar wide in-degree distributions, rapidly decaying out-degree distributions, and similar sizes. In both networks, $IN$ is larger than $OUT$. This is especially apparent in Twitter. An interesting observation is that the susceptibilities in the WWW are, especially near their maximum, much higher than those of Twitter. The difference is a striking two orders of magnitude! We suggest that such high susceptibilities in the WWW are due to the highly modular structure of the network. This modular structure results in very large sizes of the in- and out-components of vertices either in  $IN$, $OUT$ or $F$ in the undamaged samples of the WWW network.
More specifically, the undamaged WWW has $12874$ nontrivial strongly connected components (SCCs), i.e., ones of size at least 2. There are $19$ SCCs with sizes greater than $100$. The size of the second largest SCC is $968$. The mean size of SCCs, excluding the giant (largest) SCC, is $6.37$. In contrast, the Twitter sample has only $3854$ nontrivial SCCs, none of which are larger than $100$ vertices. The second largest SCC has only $17$ vertices. The mean size of SCCs , excluding the giant (largest) SCC, is $2.23$.
The susceptibilities $\chi_d$ and $\chi_{out}^{(S)}$ of the Twitter network in Fig. \ref{fig:real2} (b) demonstrate non-monotonous behavior as a function of $p$ similar to the one in the \emph{C. elegans} network. Apart from the peak at $p_c$ there is a peak at $p=1$ in contrast to   $\chi_{in}^{(S)}$, which decreases monotonously at $p > p_c$. The origin of this behavior is unclear.
%Comparing  the susceptibility of $OUT$ in WWW and Twitter we made an interesting observation that the susceptibility of the $OUT$ in the WWW are, especially near their maximum, much higher than those of Twitter. The difference is a striking two orders of magnitude! We suggest that such high susceptibilities in the WWW are due to the highly modular structure of the network.
Further investigations are required to explain our numerical findings in detail, but based on the results presented above one can see how a simple and straightforward analysis of the susceptibility of directed networks can reveal their structural peculiarities, such as those found in the \emph{C. elegans}, WWW, and Twitter.

\section{Conclusion}
\label{conclusion}

In this paper, we studied the sensitivity of directed networks with both directed and bidirectional edges to the addition and pruning of edges and vertices. We demonstrated that different network parts [the giant strongly connected component $G_S$, which is a central core of the  network, the sets $IN$ and $OUT$ playing the role of the incoming and outgoing terminals for $G_S$, and the finite components $F$ including tendrils, tubes, and disconnected finite clusters (see Fig. \ref{fig:structure})] have different sensitivities to the addition and pruning of edges and vertices since these parts have different topological properties. It is not surprising that the sensitivities of the network parts to the addition and pruning of edges and vertices are quantified by different susceptibilities. We introduced the susceptibilities, using a relation between the percolation problem and the Potts model. For this purpose we introduced the two-point connectivity function, which characterizes whether any two vertices are connected by a directed path or not. This two-point connectivity function allowed us to find explicitly the susceptibilities of the network parts. Since it is computationally inefficient to find this function  in a large network, we also proposed an alternative method for finding the susceptibilities. Our method is based on the fact that, according to Eqs. (\ref{eq: 9}),  (\ref{eq: 25}) and (\ref{eq: 26}), the susceptibilities are determined explicitly by the mean size of the finite individual in- and out- components of vertices in the corresponding network parts, i.e., $IN$, $OUT$, and $F$. We found analytically the susceptibilities in directed uncorrelated random networks by use of the generating function method. We showed that the susceptibilities diverge at the critical point of the directed percolation transition, signaling the appearance (or disappearance) of the giant strongly connected component in the infinite size limit. In finite networks due to the finite size effect, the susceptibilities demonstrate a sharp peak  at the percolation point.
We performed numerically the statistical analysis of the individual in- and out-components of vertices and found the susceptibilities of randomly damaged real and synthetic directed complex networks, such as the World Wide Web,  Twitter, the neural network of \emph{Caenorhabditis elegans}, the Gnutella p2p filesharing network, and directed Erd\H os-R\'enyi graphs. Our analysis revealed a non-monotonous dependence of the sensitivity of $OUT$  to random pruning of edges or vertices in \emph{Caenorhabditis elegans}  and Twitter. This behavior manifests  specific structural peculiarities of these networks. Our preliminary analysis pointed out the possible role of chain-like motives in the observed effects. Comparing  the susceptibility of $OUT$ in the WWW and Twitter
%we made the interesting observation that the susceptibility of $OUT$ in the WWW and Twitter,
we made the interesting observation that the former, especially near their maximum, is two orders of magnitude higher than the latter.
We suggest that such high susceptibility of the WWW is due to the modular structure of the network. Further investigations are necessary to explain our numerical findings in detail.
We believe that measurements of the sensitivity of different parts of directed networks to the addition or pruning of edges and vertices can be an effective method for studying structural peculiarities of the networks.

% -------------------------- Acknowledgements ----------------------------------------------------------------------------
\section{Acknowledgements}
This work was supported by
%the FET proactive IP project MULTIPLEX 317532 and
the grant PEST UID/CTM/50025/2013.

\appendix

\section{Generating function technique}
\label{generating functions}

Let us consider directed  uncorrelated random complex networks.
On-site correlations between in- and out-degrees are characterized by a function  $P(q_i,q_o)$, which is the probability that a randomly chosen vertex has in-degree $q_i$ and out-degree $q_o$. These complex network have locally tree-like structure that enables us to use the generating function technique \cite{newman2001random,dorogovtsev2001giant} in order to find the size of the giant strongly connected component and statistics of finite directed components. We consider the case of a randomly damaged network and  $p$ is  the occupation probability of vertices in the considered network.

Let us first consider the following process in a graph $\mathcal{G}$. Choose at random an edge and move along this edge  forwards.
%The process is stopped when encountering either vertices with no outgoing edges or removed vertices.
We define  $Y_{out}(s)$ as the probability to reach $s$ vertices by following edges forwards. We also define the probability $Y_{in}(s)$ to reach $s$ vertices by following edges backwards.
%(in this case, the process is stopped when encountering either vertices with no incoming edges or removed vertices).
We introduce generating functions,
 \begin{eqnarray}
  H_{out}(x)&=& \sum_{s=0}^{\infty} x^s Y_{out}(s),
  \nonumber \\
   H_{in}(x) &=& \sum_{s=0}^{\infty} x^s Y_{in}(s),
  \label{eq: A0}
  \end{eqnarray}
%These functions
%$H_{in}(x)$ and $H_{out}(x)$
which are determined by the following self-consistency equations \cite{newman2001random,dorogovtsev2001giant}:
\begin{eqnarray}
H_{out}(x) &{=}&1{-}p {+}\frac{px}{\langle q_i \rangle}\sum_{q_i, q_{o}} q_i P(q_i,q_o) [H_{out}(x)]^{q_o},
\label{eq: A1} \\
H_{in}(x) &{=}&1{-}p {+}\frac{px}{\langle q_o \rangle}\sum_{q_i, q_{o}} q_o P(q_i,q_o) [H_{in}(x)]^{q_i} .
  \label{eq: A2}
\end{eqnarray}
%A diagram representation is shown in Fig. \ref{}.
There is a critical point
\begin{equation}
  p_c=\frac{\langle q_o \rangle}{\langle q_i q_o \rangle}
  \label{eq: A15}
\end{equation}
below which, i.e., at $p<p_c$, equations (\ref{eq: A1}) and (\ref{eq: A2}) have the only one solution corresponding to $H_{in}(1)=H_{out}(1)=1$. At $p > p_c$, another solution corresponding to $H_{in}(1)\equiv x_c <1$ and $H_{out}(1)\equiv y_c <1$ appears.
%The parameters   $x_c$ and $y_c$ have the following physical meaning:
$y_c$ is the probability that choosing an edge at random and moving along its direction we will reach a finite number of vertices while $x_c$ is the probability that choosing an edge at random and moving against its direction we will reach a finite number of vertices.

%In a randomly damaged network with the occupation probability $p$,
The total number of remaining vertices in the damaged network is $Np$. Let us define parameters $S_S$, $S_{IN}$, $S_{OUT}$, and $S_{F}$ as the probabilities that a vertex chosen at random among the remaining $Np$ vertices belongs to $G_S$, $IN$, $OUT$, and $F$, respectively.
%Therefore,  $Np S_S$ vertices belong to the giant strongly connected component.
The sizes of $G_S$, $IN$, $OUT$, and $F$ are $Np S_S$, $Np S_{IN}$, $Np S_{OUT}$, and $Np S_{F}$, respectively.
In turn, the fraction $S_{S}$ of vertices belonging to $G_{S}$  ($S_S = N(G_{S})/ Np$) is the probability that a randomly chosen vertex has at least one in-coming edge from $G_S$ and at least one outgoing edge leading to $G_{S}$. Using this relation, we find
\begin{equation}
  S_{S} = \sum_{q_i, q_{o}} (1-x_{c}^{q_i})(1-y_{c}^{q_o}) P(q_i,q_o).
 \label{eq: A8a}
\end{equation}
The fraction $S_{IN}$ of vertices belonging $IN$  is the probability that a randomly chosen vertex has no in-coming edge from $G_{S}$ but at least one outgoing edge leading to  $G_{S}$:
\begin{equation}
  S_{IN} = \sum_{q_i, q_{o}} x_{c}^{q_i}(1-y_{c}^{q_o}) P(q_i,q_o).
  \label{eq: A8b}
\end{equation}
The fraction $S_{OUT}$ is the probability that a randomly chosen vertex has at least one  incoming edge from $G_{S}$ but no outgoing edge leading to $G_{S}$
\begin{equation}
  S_{OUT} = \sum_{q_i, q_{o}} (1-x_{c}^{q_i}) y_{c}^{q_o} P(q_i,q_o).
 \label{eq: A8c}
\end{equation}
The fraction $S_{F}$ of vertices belonging to $F$ is the probability that a vertex chosen at random has no incoming and no outgoing edges coming from or leading to $G_S$,
\begin{equation}
  S_{F} = \sum_{q_i, q_{o}} x_{c}^{q_i}y_{c}^{q_o} P(q_i,q_o).
  \label{eq: A8d}
\end{equation}
Introducing a generating function
\begin{equation}
  \Phi(x,y) \equiv \sum_{q_i, q_o} x^{q_i} y^{q_o} P(q_i, q_o),
  \label{eq: 30}
\end{equation}
%Using the generating function $\Phi (x, y)$ given by Eq. (\ref{eq: 30}),
we can write the fractions of vertices in  $G_{S}$, $IN$, $OUT$, $F$, and $G_{IO}$ in the following form,
\begin{eqnarray}
S_{S}&=&1 - \Phi (x_c, 1) - \Phi (1, y_c) + \Phi (x_c, y_c),
\label{eq: A3} \\
S_{IN}&=& \Phi (x_c, 1) -  \Phi (x_c, y_c),
\label{eq: A4} \\
S_{OUT} &=&  \Phi (1, y_c) - \Phi (x_c, y_c) ,
\label{eq: A5} \\
S_{F} &=& \Phi (x_c, y_c),
\label{eq: A6} \\
S_{IO} &=&  S_{S}+S_{IN}+S_{OUT} = 1 - S_{F},
\label{eq: A7} \\
1 &=&  S_{S}+S_{IN}+S_{OUT}+S_{F},
\label{eq: A8}
\end{eqnarray}
%Note that in the case of a randomly damaged network with the occupation probability $p$, the total number of remaining vertices in the network is $Np$. Therefore,  $Np S_S$ vertices belong to the giant strongly connected component. The sizes of $IN$, $OUT$, and $F$ are $Np S_{IN}$, $Np S_{OUT}$, and $Np S_{F}$,respectively.
%The size of $F$ is $Np S_{F}$.
%Therefore, $S_S$, $S_{IN}$, $S_{OUT}$, and $S_{F}$ are the probabilities that a vertex chosen at random among the remaining vertices belongs to $G_S$, $IN$, $OUT$, and $F$, respectively, in the damaged network.

Solving Eqs. (\ref{eq: A1}) at $0 < p -p_c \ll p_c$ in the leading order in $(p -p_c)/ p_c$, we find
\begin{eqnarray}
x_c \approx 1- \frac{2 \langle q_o\rangle (p -p_c)}{p_{c}^2 \langle q_o q_i (q_i -1) \rangle},
\label{eq: A9} \\
y_c \approx 1- \frac{2 \langle q_i\rangle (p -p_c)}{p_{c}^2 \langle q_o q_i (q_o -1) \rangle}.
  \label{eq: A10}
\end{eqnarray}
Substituting this result into Eqs. (\ref{eq: A3})--(\ref{eq: A7}) gives the critical behavior
\begin{eqnarray}
S_{S}\propto \Bigl(\frac{p -p_c}{p_c}\Bigr)^2,
\label{eq: A11} \\
S_{IO}, S_{IN}, S_{OUT} \propto \frac{p -p_c}{p_c},
\label{eq: A12} \\
S_{F} \approx 1 - O\Bigl(\frac{p -p_c}{p_c}\Bigr).
\label{eq: A13}
\end{eqnarray}
The derivatives $d H_{out}(x)/dx|_{x=1}$ and $d H_{in}(x)/dx|_{x=1}$
can be found from Eqs. (\ref{eq: A1}) and (\ref{eq: A2}):
\begin{eqnarray}
\frac{d H_{out}(x)}{dx}\Big|_{x=1}=\frac{\frac{p}{\langle q_i \rangle}\sum_{q_i, q_{o}} q_i P(q_i,q_o) y^{q_o}_c }{1-\frac{p}{\langle q_i \rangle}\sum_{q_i, q_{o}} q_i q_o P(q_i,q_o)  y^{q_o-1}_c},
\nonumber \\
\frac{d H_{in}(x)}{dx}\Big|_{x=1}=\frac{\frac{p}{\langle q_o \rangle}\sum_{q_i, q_{o}}  q_o P(q_i,q_o)x^{q_i}_c}{1-\frac{p}{\langle q_o \rangle}\sum_{q_i, q_{o}} q_i q_o P(q_i,q_o)  x^{q_i-1}_c} .
\nonumber \\
\label{eq: A14}
\end{eqnarray}
These derivatives are positive both below and above $p_c$. They diverge at the critical point
$p=p_c$. At $p < p_c$ we find the explicit result for the derivatives Eq. (\ref{eq: A14}):
\begin{equation}
  \frac{d H_{out}(x)}{dx}\Big|_{x=1} = \frac{d H_{in}(x)}{dx}\Big|_{x=1} = \frac{p p_{c}}{p_c -p}
  \label{eq: A16 below}
\end{equation}
At $p > p_c$, substituting Eqs. (\ref{eq: A9}) and (\ref{eq: A10}) into Eq. (\ref{eq: A14}), we obtain  the following critical behavior in the leading order in $ (p -p_c)/ p_c \ll 1$:
\begin{equation}
  \frac{d H_{out}(x)}{dx}\Big|_{x=1} \approx \frac{d H_{in}(x)}{dx}\Big|_{x=1} \approx \frac{p p_{c}}{p -p_c}
  \label{eq: A16}
\end{equation}

\section{Generating functions for finite components above $p_c$}
\label{generating functions: finite components}

%%%%%%%%%%%%%%%%%%%%%%%%%%%%%%%%%%%%%%%%%%%%%%%%%%%%%%%%%%%%%%%%%%%%%%%%%%%%move to App
The generating functions $H_{out}(x)$ and $H_{in}(x)$, Eqs. (\ref{eq: A1}) and (\ref{eq: A2}), do not allow us to find statistics of individual finite in- and out-components of vertices in the finite component $F$ above the percolation threshold.  For this purpose we introduce other generating functions as  follows.
Choose at random an edge and move along this edge forwards. We define  $Y_{out}^{(F)}(s)$ as the probability to reach $s$ vertices, which have no incoming edges
%going from $G_S$.
by which one can reach $G_S$, moving backwards.
Then, choose at random an edge and move backwards. We define $Y_{in}^{(F)}(s)$ as the probability to reach $s$ vertices (moving backwards), which have no outgoing edges
%leading to $G_S$,
by which one can reach $G_S$, moving forwards.
%when we move against the edge directions.
These probabilities determine the following generating functions:
 \begin{eqnarray}
  \widetilde{H}_{out}(x)&=& \sum_{s=0}^{\infty} x^s Y_{out}^{(F)}(s),
  \nonumber \\
   \widetilde{H}_{in}(x) &=& \sum_{s=0}^{\infty} x^s Y_{in}^{(F)}(s).
  \label{eq: 28}
  \end{eqnarray}
Note that $\widetilde{H}_{in}(1)$ and $\widetilde{H}_{out}(1)$
are the probabilities to reach a finite number of vertices, which have no outgoing or incoming edges from $G_S$, when we go against or along the edge directions, respectively.
We find that in uncorrelated random directed complex networks, the generating functions $\widetilde{H}_{out}(x)$ and $\widetilde{H}_{in}(x)$ are determined by the following self-consistency equations:
\begin{eqnarray}
\widetilde{H}_{out}(x) &=& 1{-}p {+}p \sum_{q_i, q_{o}} \frac{q_i}{\langle q_i \rangle}(1-x_{c}^{q_i -1}) P(q_i,q_o)y_{c}^{q_o} +
\nonumber \\
& & px\sum_{q_i, q_{o}} \frac{q_i}{\langle q_i \rangle} x_{c}^{q_i -1}P(q_i,q_o) [\widetilde{H}_{out}(x)]^{q_o},
\label{eq: A17} \\
\widetilde{H}_{in}(x) &=& 1{-}p {+}p \sum_{q_i, q_{o}} x_{c}^{q_i} P(q_i,q_o)\frac{q_o}{\langle q_o \rangle} (1-y_{c}^{q_o-1}) +
\nonumber \\
& & px\sum_{q_i, q_{o}} [\widetilde{H}_{in}(x)]^{q_i} P(q_i,q_o) \frac{q_o}{\langle q_o \rangle} y_{c}^{q_o -1},
\label{eq: A18}
\end{eqnarray}
where the parameter $x_c$ and $y_c$ are determined by Eqs. (\ref{eq: A1}) and (\ref{eq: A2}). The first term $1-p$ in Eqs. (\ref{eq: A17}) and (\ref{eq: A18}) is the probability that a vertex at the end of an edge, along which we move, is removed.
The second term is the probability that a vertex at the end of an edge, along which we move,
has at least one incoming edge
%from $G_S$
by which one can reach $G_S$, moving backwards
among $q_i -1$ incoming edges (note that one more incoming edge is the edge along which we arrive at this vertex).
The second term in Eq. (\ref{eq: A18}) is
%the probability that a vertex at the end of an edge, against of which direction we move,
the probability that a vertex at the end of an edge, along which we arrived moving backwards, has at least one outgoing edge
%leading to $G_S$
by which one can reach $G_S$, moving forwards
among $q_o -1$ outgoing edges (note that one more outgoing edge is the edge along which we arrived moving backwards).
At $x=1$ we have a solution $\widetilde{H}_{out}(1)=y_c$ and $\widetilde{H}_{in}(1)=x_c$. At $p < p_c$, we have $x_c=y_c=1$ and Eqs. (\ref{eq: A17}) and (\ref{eq: A18}) are reduced to Eqs. (\ref{eq: A1}) and (\ref{eq: A2}).
It is easy to show that the probabilities  $Y_{out}^{(F)}(s)$ and  $Y_{in}^{(F)}(s)$ defined above, are related with the solution of Eqs. (\ref{eq: A17}) and (\ref{eq: A18}) as follows:
\begin{eqnarray}
  \frac{d^s \widetilde{H}_{out}(x)}{s! d^s x}\Big|_{x=0} = Y_{out}^{(F)}(s),
  \nonumber \\
  \frac{d^s \widetilde{H}_{in}(x)}{s! d^s x}\Big|_{x=0} = Y_{in}^{(F)}(s).
  \label{eq: 28b}
  \end{eqnarray}
The first derivatives of $\widetilde{H}_{out}(x)$ and $\widetilde{H}_{in}(x)$ at $x=1$,
\begin{eqnarray}
  d \widetilde{H}_{out}(x) / dx |_{x=1} = \sum_{s=0}^{\infty} s Y_{out}^{(F)}(s),
   \nonumber \\
   d \widetilde{H}_{in}(x)/ dx |_{x=1} = \sum_{s=0}^{\infty} s Y_{in}^{(F)}(s),
   \label{eq: 29}
  \end{eqnarray}
give the mean number of vertices, which are reachable by following edges either forwards or backwards and which have either no incoming or no outgoing edges with $G_S$, respectively.
Differentiating Eqs. (\ref{eq: A17}) and (\ref{eq: A18}) with respect to $x$, we find
\begin{eqnarray}
\frac{d \widetilde{H}_{out}(x)}{dx}\Big|_{x{=}1}&{=}&\frac{\frac{p}{\langle q_i \rangle}\sum_{q_i, q_{o}} q_i P(q_i,q_o) x^{q_i-1}_c y^{q_o}_c }{1{-}\frac{p}{\langle q_i \rangle}\sum_{q_i, q_{o}} q_i q_o P(q_i,q_o) x^{q_i-1}_c y^{q_o-1}_c},
\nonumber \\
\frac{d \widetilde{H}_{in}(x)}{dx}\Big|_{x{=}1}&{=}&\frac{\frac{p}{\langle q_o \rangle}\sum_{q_i, q_{o}}  q_o P(q_i,q_o)x^{q_i}_c y^{q_o-1}_c}{1{-}\frac{p}{\langle q_o \rangle}\sum_{q_i, q_{o}} q_i q_o P(q_i,q_o)  x^{q_i-1}_c y^{q_o-1}_c} .
\nonumber \\
\label{eq: A19}
\end{eqnarray}
Using Eqs. (\ref{eq: A9}) and (\ref{eq: A10}), we find that these derivatives diverge when $p$ tends to $p_c$ from above:
\begin{equation}
  \frac{d \widetilde{H}_{out}(x)}{dx}\Big|_{x=1} \approx \frac{d \widetilde{H}_{in}(x)}{dx}\Big|_{x=1} \approx \frac{p p_{c}}{3(p-p_c)},
  \label{eq: A20}
\end{equation}

The individual in- and out-components of vertex in $F$ are the sum of the number of vertices reachable by following $q_i$ and $q_o$ edges backwards or forwards, respectively, but intersections with $IN$ and $OUT$ must be excluded [see Eq. (\ref{eq: 11})]. The mean values of the individual in- and out-components can be found by use of the functions $\widetilde{H}_{out}(x)$, $\widetilde{H}_{out}(x)$, and   $\Phi(x,y)$,
\begin{eqnarray}
\langle s_{out}^{(F)} \rangle &{=}&  \frac{d \Phi(x_c, \widetilde{H}_{out}(x))}{S_F d x}\Big|_{x=1}
\nonumber \\
&{=}& \sum_{q_i, q_{o}} P(q_i,q_o) q_o x^{q_i}_c y^{q_o {-}1}_c  \frac{d \widetilde{H}_{out}(x)}{S_F dx}\Big|_{x=1},
\label{eq: 31} \\
\langle s_{in}^{(F)} \rangle &{=}&  \frac{d \Phi(\widetilde{H}_{in}(x)),y_c)}{S_F d x}\Big|_{x=1}
\nonumber \\
& {=}& \sum_{q_i, q_{o}}  P(q_i,q_o) q_i x^{q_i{-}1}_c y^{q_o}_c \frac{d \widetilde{H}_{in}(x)}{S_F dx}\Big|_{x=1}.
\label{eq: 32}
\end{eqnarray}
Here the derivatives are given by Eq. (\ref{eq: A19}).  At $p < p_c$ we have $x_c = y_c =1$ and  $S_F =1$ since $F=\mathcal{G}$.

The mean size of the individual in-component of vertices belonging to $IN$ and the mean size of the individual out-component of vertices belonging to $OUT$ can be found by use of the generating functions $H_{in}(x)$ and $H_{out}(x)$ [see Eqs. (\ref{eq: A1}) and (\ref{eq: A2})], respectively.
We replace $x_c$  to $H_{in}(x)$ in Eq. (\ref{eq: A4}) and $y_c$ to $H_{out}(x)$ in (\ref{eq: A5}). Differentiating  the obtained functions with respect to $x$, we find
\begin{eqnarray}
\!\!\!\!\!\!\!\!\!\!\!\!\!\!\!\!\!\!\!\langle s_{out}^{(OUT)} \rangle &{=}& \frac{d[\Phi (1, H_{out}(x)) {-}  \Phi (x_c, H_{out}(x))]}{ S_{OUT} dx} \Big|_{x{=}1},
\label{eq: 36} \\
\langle s_{in}^{(IN)} \rangle &{=}&
\frac{d[\Phi (H_{in}(x), 1) {-}  \Phi (H_{in}(x), y_c)]}{S_{IN} dx} \Big|_{x{=}1}.
%\nonumber \\
%&=&
%\sum_{q_i, q_{o}}  P(q_i,q_o) q_i x^{q_i-1}_c (1- y^{q_o}_c )\frac{d H^{(i)}(x)}{dx}\Big|_{x=1}.
\label{eq: 37}
\end{eqnarray}

\section{Susceptibility of networks with  directed and bidirectional edges below $p_c$}
\label{susc bidirectional}

The percolation transition in random complex networks with directed and bidirectional edges was studied in \cite{boguna2005generalized}. These networks are described by the probability $P(q_i, q_o, q_b)$ that a vertex has $q_i$ incoming edges, $q_o$ outgoing edges, and $q_o$ bidirectional edges. This joint degree distribution must satisfy the condition $\langle q_i \rangle = \langle q_o \rangle$. In this section we use the generating function technique to find the susceptibilities $\chi_d$ and $\chi_t$ [Eqs. (\ref{eq: 9}) and (\ref{eq: 10})] in  uncorrelated random directed networks with directed and bidirectional edges.

Let us consider randomly damaged network and $p$ is the occupation probability. We will only consider the case $p < p_c$ for simplicity.
At $p < p_c$ we introduce three generating functions $H_{in}(x)$, $H_{out}(x)$, and $H_{b}(x)$. They are determined by the following equations:
\begin{eqnarray}
H_{out}(x) &=& 1{-}p {+}\frac{px}{\langle q_i \rangle}\sum_{q_i, q_{o}, q_b} q_i P(q_i,q_o,q_b) \times
\nonumber \\
& &[H_{out}(x)]^{q_o} [H_{b}(x)]^{q_b},
\nonumber \\
H_{in}(x) &=& 1{-}p {+}\frac{px}{\langle q_o \rangle}\sum_{q_i, q_{o}, q_b} q_o P(q_i,q_o, q_b)\times
\nonumber \\
& & [H_{in}(x)]^{q_i} [H_{b}(x)]^{q_b},
\nonumber \\
H_{b}(x) &=& 1{-}p {+}\frac{px}{\langle q_b \rangle}\sum_{q_i, q_{o}, q_b} q_b P(q_i,q_o,q_b) \times
\nonumber \\
& &[H_{out}(x)]^{q_o} [H_{b}(x)]^{q_b -1},
\label{eq: A21}
\end{eqnarray}
These equations have a solution with $H_{in}(1)=H_{out}(1)=H_{b}(1)=1$ at $p < p_c$.
 %that means that all vertices have finite in- and out-components.
A solution with $H_{in}(1), H_{out}(1), H_{b}(1) < 1$ appears at $p > p_c$. The critical point $p_c$ is given by a quadratic equation:
\begin{eqnarray}
 0 &{=}& p_{c}^2 \Bigr[\langle q_i q_o \rangle  \langle q_b (q_b {-}1) \rangle {-}\langle q_i q_b \rangle  \langle q_o q_b \rangle \Bigl] {-}
  \nonumber \\
   & & p_{c} \Bigr[\langle q_i q_o \rangle  \langle q_b \rangle {+} \langle q_b (q_b {-}1) \rangle  \langle q_i \rangle \Bigl] {+} \langle q_i \rangle \langle q_b \rangle.
  \label{eq: A22}
\end{eqnarray}
%%%%%%%%%%%%%%%%%%%%%%%%%%%%%%%%%%%%%%%%%%%%%%%%%%%%%%%%%%%%%%%%%%%%%%%%%%%%%%%%%%
The first derivatives of the functions $H_{in}(x)$, $H_{out}(x)$, and $H_{b}(x)$ at $x=1$ diverge at the critical point $p=p_c$:
\begin{equation}
  \frac{d H_{out}(x)}{dx}\Big|_{x=1}, \frac{d H_{out}(x)}{dx}\Big|_{x=1}, \frac{d H_{b}(x)}{dx}\Big|_{x=1}  \propto  \frac{p}{(p_c -p)}.
  \label{eq: A23}
\end{equation}
%Therefore, the mean sizes of the in- and out-components of vertices diverge as $p_c /|p-p_c|$ when $p$ tends to $p_c$.

Introducing the generating function
\begin{equation}
  \Phi(x,y,z) \equiv \sum_{q_i, q_o, q_b} x^{q_i} y^{q_o} z^{q_b} P(q_i, q_o, q_b),
  \label{eq: 41}
\end{equation}
we find the mean sizes of the in- and out-components of vertices at $p \leq p_c$
\begin{eqnarray}
\langle s_{out} \rangle &=& \frac{d \Phi(1,H_{out}(x),H_{b}(x))}{d x}\Big|_{x=1}
\label{eq: 42} \\
\langle s_{in} \rangle &=&  \frac{d \Phi(H_{in}(x),1,H_{b}(x)}{d x}\Big|_{x=1}
\label{eq: 43}
\end{eqnarray}
Using Eq. (\ref{eq: A23}), we find that $\langle s_{out} \rangle$ and $\langle s_{in} \rangle$  diverge as $p_c /|p-p_c|$ when $p$ tends to $p_c$.
Therefore, the susceptibilities $\chi_d$ and $\chi_t$ also diverge, $\chi_d , \chi_t \propto 1/(p-p_c)$, signaling the percolation phase transition.
% -------------------------- Bibliography
%--------------------------------------------------------------------------------
\bibliography{mybib_susceptibility}
\end{document}